\documentclass[twoside,twocolumn,9pt]{article}
\usepackage{extsizes}
\usepackage[super,sort&compress,comma]{natbib} 
\usepackage[version=3]{mhchem}
\usepackage[left=1.5cm, right=1.5cm, top=1.785cm, bottom=2.0cm]{geometry}
\usepackage{balance}
\usepackage{times,mathptmx}
\usepackage{sectsty}
\usepackage{graphicx} 
\usepackage{lastpage}
\usepackage[format=plain,justification=justified,singlelinecheck=false,font={stretch=1.125,small,sf},labelfont=bf,labelsep=space]{caption}
\usepackage{float}
\usepackage{fancyhdr}
\usepackage{fnpos}
\usepackage[english]{babel}
\addto{\captionsenglish}{%
  \renewcommand{\refname}{References}
}
\usepackage{array}
\usepackage{droidsans}
\usepackage{charter}
\usepackage[T1]{fontenc}
\usepackage[usenames,dvipsnames]{xcolor}
\usepackage{setspace}
\usepackage[compact]{titlesec}
\usepackage{hyperref}

\usepackage{amsmath}
\usepackage{amsfonts}
\usepackage{amssymb}
\usepackage{bm}

\newcommand*{\citen}[1]{%
  \begingroup
    \romannumeral-`\x 
    \setcitestyle{numbers}%
    \cite{#1}%
  \endgroup
}

\DeclareMathAlphabet\mathzapf       {T1}{pzc} {mb} {it}

\definecolor{cream}{RGB}{222,217,201}

\begin{document}

\pagestyle{plain}
\thispagestyle{plain}

\makeFNbottom
\makeatletter
\renewcommand\LARGE{\@setfontsize\LARGE{15pt}{17}}
\renewcommand\Large{\@setfontsize\Large{12pt}{14}}
\renewcommand\large{\@setfontsize\large{10pt}{12}}
\renewcommand\footnotesize{\@setfontsize\footnotesize{7pt}{10}}
\makeatother

\renewcommand{\thefootnote}{\fnsymbol{footnote}}
\renewcommand\footnoterule{\vspace*{1pt}%
\color{cream}\hrule width 3.5in height 0.4pt \color{black}\vspace*{5pt}} 
\setcounter{secnumdepth}{5}

\makeatletter 
\renewcommand\@biblabel[1]{#1}
\renewcommand\@makefntext[1]%
{\noindent\makebox[0pt][r]{\@thefnmark\,}#1}
\makeatother 
\renewcommand{\figurename}{\small{Fig.}~}
\sectionfont{\sffamily\Large}
\subsectionfont{\normalsize}
\subsubsectionfont{\bf}
\setstretch{1.125} 
\setlength{\skip\footins}{0.8cm}
\setlength{\footnotesep}{0.25cm}
\setlength{\jot}{10pt}
\titlespacing*{\section}{0pt}{4pt}{4pt}
\titlespacing*{\subsection}{0pt}{15pt}{1pt}

\fancyfoot{}
\fancyfoot[RO]{\footnotesize{\sffamily{1--\pageref{LastPage} ~\textbar  \hspace{2pt}\thepage}}}
\fancyfoot[LE]{\footnotesize{\sffamily{\thepage~\textbar\hspace{3.45cm} 1--\pageref{LastPage}}}}
\fancyhead{}
\renewcommand{\headrulewidth}{0pt} 
\renewcommand{\footrulewidth}{0pt}
\setlength{\arrayrulewidth}{1pt}
\setlength{\columnsep}{6.5mm}
\setlength\bibsep{1pt}

\makeatletter 
\newlength{\figrulesep} 
\setlength{\figrulesep}{0.5\textfloatsep} 

\newcommand{\topfigrule}{\vspace*{-1pt}%
\noindent{\color{cream}\rule[-\figrulesep]{\columnwidth}{1.5pt}} }

\newcommand{\botfigrule}{\vspace*{-2pt}%
\noindent{\color{cream}\rule[\figrulesep]{\columnwidth}{1.5pt}} }

\newcommand{\dblfigrule}{\vspace*{-1pt}%
\noindent{\color{cream}\rule[-\figrulesep]{\textwidth}{1.5pt}} }

\makeatother

\twocolumn[
  \begin{@twocolumnfalse}
  {\fontsize{14.5}{0} \selectfont \textbf{Dynamical self-assembly of dipolar active Brownian particles in two dimensions}} \\
  \\
  \large{Guo-Jun Liao,$^{\ast}$\textit{$^{a}$} Carol K. Hall,\textit{$^{b}$} and Sabine H. L. Klapp$^{\ast}$\textit{$^{a}$}} \\
  \\
  \normalsize{%
Based on Brownian Dynamics (BD) simulations, we study the dynamical self-assembly of active Brownian particles with dipole-dipole interactions, stemming from a permanent point dipole at the particle center. 
The propulsion direction of each particle is chosen to be parallel to its dipole moment. 
We explore a wide range of motilities and dipolar coupling strengths and characterize the corresponding behavior based on several order parameters.
At low densities and low motilities, the most important structural phenomenon is the aggregation of the dipolar particles into chains.
Upon increasing the particle motility, these chain-like structures break, and the system transforms into a weakly correlated isotropic fluid.
At high densities, we observe that the motility-induced phase separation is strongly suppressed by the dipolar coupling.
Once the dipolar coupling dominates the thermal energy, the phase separation disappears, and the system rather displays a flocking state, where particles form giant clusters and move collective along one direction.
We provide arguments for the emergence of the flocking behavior, which is absent in the passive dipolar system.
} \\
 \end{@twocolumnfalse} \vspace{0.6cm}

  ]

\renewcommand*\rmdefault{bch}\normalfont\upshape
\rmfamily
\section*{}
\vspace{-1cm}


\footnotetext{\textit{$^{a}$~
Institut f\"ur Theoretische Physik, Technische Universit\"at Berlin, Hardenbergstr. 36, D-10623 Berlin, Germany. E-mail: guo-jun.liao@campus.tu-berlin.de, klapp@physik.tu-berlin.de}}
\footnotetext{\textit{$^{b}$~Department of Chemical \& Biomolecular Engineering, North Carolina State
University, Raleigh, NC 27695, USA. }}



\section{\label{sec:Intro}Introduction}

%
Self-propelled particles are capable of converting energy from an internal source or the surroundings into their own active motion. \cite{Romanczuk2012, Zottl2016}
It is now well established that active motion of individual particles, combined with different types of interactions among these particles, can lead to remarkable collective behavior.
For example, repulsive interactions alone can generate the so-called motility-induced phase separation, where large, densely packed clusters coexist with freely moving particles. \cite{Buttinoni2013}
A standard model describing this non-equilibrium phase transition is that of active Brownian particles, where each particle moves with a constant propulsion speed along a direction subject to white noise, and the interactions are purely steric and isotropic.
Clearly, one could expect more complex behavior in active systems with anisotropic interactions.
This situation has been investigated by a large number of theoretical and simulation studies. \cite{Toner1995, Bertin2006, Chate2008, Solon2015}
A famous example in this direction is the Vicsek model, where active, point-like particles interact such that they tend to align their velocities with those of their neighbors.
The resulting collective behavior includes traveling bands \cite{Chate2008} and flocking. \cite{Vicsek1995}
A further representative system is a suspension of active Brownian particles with polar interactions favoring parallel orientations of the velocities irrespective of the spatial configuration.
This system displays a state with moving lanes and bands even at low densities, where conventional active Brownian particles do not show significant behavior. \cite{Martin-Gomez2018}
Further, at high densities, the polar coupling favors motility-induced phase separation compared to the non-polar case. \cite{Sese-Sansa2018}
%

%
In most studies of active systems with anisotropic interactions, they are assumed to be of short range. \cite{Bechinger2016}
Less effort has been spent on active systems involving long-range anisotropic interactions, such as dipole-dipole interactions stemming from intrinsic (permanent) dipole moments or dipoles induced by an electric or magnetic field. \cite{Kogler2015} 
A notable feature of dipolar interactions is that they depend not only on the orientations of the two involved particles, but also on their configuration in space.
Specifically, two particles with point dipoles at their center tend to align head-to-tail, where the arrowhead of one dipole moment is directed toward the arrowtail of the other dipole. 
In contrast, particles configured side by side favor antiparallel alignment.
Due to this complexity, one would expect a system of dipolar active particles to show macroscopic structures significantly different from active models with simple polar interactions, such as the Vicsek model \cite{Vicsek1995} and its variants.
%

%
There are some recent studies, which have focused particularly on the dynamics of dipolar active colloids.
For example, ref. \citen{Kaiser2015} and \citen{Guzman-Lastra2016} have studied the structural transformation of small dipolar clusters under the impact of activity starting from different (meta)stable configurations, either neglecting \cite{Kaiser2015} or including \cite{Guzman-Lastra2016} hydrodynamic interactions. 
Another example is a system of spherical, ferromagnetic rollers confined to a fluctuating surface.
Here it has been shown, both, experimentally and in simulations, that these particles can exhibit swarming or vortex patterns when energized by a vertical alternating field. \cite{Kaiser2017}
Owing to the interplay between self-propulsion and dipolar interactions, these vortices persist even on a flat surface. \cite{Kokot2018}
In nature, magnetotactic bacteria are known to sense the earth's magnetic field and move along or against the field direction. \cite{Blakemore1982, Frankel1984, Klumpp2019} 
Experiments have observed that these bacteria, when confined in a microfluidic channel and placed under an external magnetic field, display clustering behavior along the channel. \cite{Waisbord2016}
The underlying mechanism of this clustering instability was also investigated analytically and in simulations. \cite{Meng2018} 
We also mention dipole-like, active Janus particles with two screened electric charges in each hemisphere.
It is found that these particles can self-organize into fingerprint-like patterns at high densities. \cite{Harder2018}
Moreover, an external electric or magnetic field can induce two point dipoles in the respective hemispheres of a Janus particle and thereby further complicate the interactions between particles.
By fine-tuning the interactions, it has been reported that the external field can be used to control the collective behavior of Janus particles. \cite{Yan2016, Han2017}
However, these latter models involve even more complex anisotropic interactions as compared to the purely dipolar case. 
Indeed, the collective behavior of large ensembles of the simplest dipolar active model, that is, dipolar active Brownian particles, is not yet explored.
This is the motivation for the present work.
%

%
Specifically, we present Brownian Dynamics simulation results for the dynamical self-assembly of a two-dimensional system of active particles, where each particle has a permanent point dipole moment oriented in the plane.
In addition, each particle is subject to a self-propulsion force, which is directed along the dipole, as well as to thermal noise. 
We investigate the collective behavior of our model system for three density regimes, considering a wide range of motilities and dipolar coupling strengths.
Our model can be realized, \textit{e.g.}, by Janus particles with a magnetic material coated on one of the hemispheres. \cite{Baraban2013, Baraban2013a}
In that case, both the dipole moment and the propulsion force of the Janus particle are directed along its symmetric axis.

%
For all three densities we present state diagrams illustrating the complex interplay between essentially three phenomena:
chain formation (which already occurs in the passive case), motility-induced phase separation, and polar ordering.

%
%
%
%
%

%
The remainder of this paper is organized as follows. 
In Sec.~\ref{sec:Method} we present our model of dipolar active particles and the simulation details, as well as the target quantities and order parameters investigated.
Based on analysis of these quantities for a range of parameters, we discuss the collective behavior of the model system in Sec.~\ref{sec:result}.
Finally, in Sec.~\ref{sec:conclusions} we summarize our findings.
\section{\label{sec:Method}Model and methods of investigation}
\subsection{\label{sec:model}Model system}
Our dipolar active system consists of $N$ disk-shaped Brownian particles with diameter $\sigma$ dispersed in a monolayer in the $xy$-plane.
Each particle carries a fluctuating point dipole moment $\boldsymbol{\mu}_i$ ($i = 1, ..., N$) located at its center.
For passive monolayers of dipolar disks, it is well established that the particles tend to orient along in-plane directions to form chains and rings, \cite{Tavares2002, Duncan2004, Duncan2006, Kantorovich2008, Cerda2008} or dense ordered states, \cite{Tavares2002, Klapp2002, Ouyang2011, Geiger2013} if the dipolar interactions are sufficiently strong.
Having this in mind, we assume beforehand that $\boldsymbol{\mu}_i$ lies in the $xy$-plane, \textit{i.e.}, fluctuations in $z$-direction are neglected.
To model the self-propulsion, we assume that each particle is subject to a force $\boldsymbol{F}_i$, which has a constant magnitude and is directed along $\boldsymbol{\mu}_i$ at each instant of time.
%

%
The pair potential between particles located at positions $\boldsymbol{r}_i$ and $\boldsymbol{r}_j$ ($i \neq j$) is of the form
\begin{equation} \label{eqn:pair}
  u_{pair}\big(\boldsymbol{r}_{ij}, 
                \boldsymbol{\mu}_i, 
                \boldsymbol{\mu}_j \big) = 
    u_{sr}\left(r_{ij}\right) + 
    u_{dd}\big(\boldsymbol{r}_{ij}, 
                \boldsymbol{\mu}_i, 
                \boldsymbol{\mu}_j \big)
      \text{,}
\end{equation}
where the first term on the right-hand side stands for the short-range steric repulsion (sr), which only depends on the distance $ r_{ij} = \vert \boldsymbol{r}_{ij} \vert = \vert \boldsymbol{r}_{j} - \boldsymbol{r}_{i} \vert$.
%
%
We assume that the steric repulsion can be described by the Weeks-Chandler-Anderson potential
\begin{equation} \label{eqn:Uwca}
u_{sr}(r_{ij})= 
  \begin{cases}
    4\epsilon \left[
      \left(\dfrac{\sigma}{r_{ij}}\right)^{12} - 
      \left(\dfrac{\sigma}{r_{ij}}\right)^{6} + 
      \dfrac{1}{4}
    \right]
    \text{,} &\text{if $r_{ij} < r_{c}$,}\\
    0\text{,} &\text{else.}
  \end{cases}
\end{equation} 
In this study, we set the length unit to be $\sigma$ and fix the repulsive strength $\epsilon^* = \beta \epsilon = 10$, where the thermal energy, $\beta^{-1} = k_{B}T$, is set to be the energy unit (with $k_B$ being Boltzmann's constant and $T$ being the temperature).
This potential is truncated at the cut-off radius $r_c = 2^{1/6} \sigma$, such that eqn~\eqref{eqn:Uwca} and its derivative vanish to zero continuously at the truncation point.
The last term in eqn~\eqref{eqn:pair} represents the (long-range) dipole-dipole interaction. 
Its functional form is given by
\begin{equation} \label{eqn:Udd}
u_{dd}\big(\boldsymbol{r}_{ij}, \boldsymbol{\mu}_{i}, \boldsymbol{\mu}_{j}\big)= 
      \dfrac{\boldsymbol{\mu}_{i} \cdot \boldsymbol{\mu}_{j}}{r_{ij}^{3}} - 
      3 \dfrac{
          \big(\boldsymbol{\mu}_{i} \cdot \boldsymbol{r}_{ij} \big)
          \big(\boldsymbol{\mu}_{j} \cdot \boldsymbol{r}_{ij} \big)
        }
      {r_{ij}^{5}}
    \text{.}
\end{equation}
We define the dipolar coupling strength as $\lambda = \beta \mu^2 \sigma^{-3}$ with $\mu$ being the magnitude of a dipole moment, \textit{i.e.}, $\boldsymbol{\mu}_i = \mu \hat{\boldsymbol{\mu}}_i$.

\subsection{\label{sec:BD}Brownian Dynamics simulation}

To investigate the system's dynamical behavior, we perform conventional Brownian Dynamics (BD) simulations without hydrodynamic interactions. The motion of the $i$th particle is then determined by the coupled Langevin equations \cite{VanTeeffelen2008} for its position $\boldsymbol{r}_i$ and orientation $\hat{\boldsymbol{e}}_i = \left( \text{cos}\psi_i, \text{sin}\psi_i \right)^T$,
\begin{align} 
  \dot{\boldsymbol{r}}_i & = 
    \beta \mathbb{D}
      \Big[
        F_{0}\widehat{\boldsymbol{e}}_{i} 
        - \nabla_{\boldsymbol{r}_i}U_{i} 
        + \boldsymbol{\xi}_{i}\left(t\right) 
      \Big], \label{eqn:coupled_Langevin_trans}\\
  \dot{\psi}_i & = 
    \beta D_r
      \Big[
        - \partial_{\psi_i} U_{i} 
        + \Gamma_{i}\left(t\right)
      \Big], \label{eqn:coupled_Langevin_rot}
\end{align}
where the dots denote time derivatives and $\psi_i$ is the polar angle. 
In Eqs.~\eqref{eqn:coupled_Langevin_trans}-\eqref{eqn:coupled_Langevin_rot}, the potential energy for the $i$th particle is given by
\begin{equation}
  U_{i} = \sum_{j = 1, j \neq i}^{N} u_{pair}\big(\boldsymbol{r}_{ij},
                                                    \boldsymbol{\mu}_{i},
                                                    \boldsymbol{\mu}_{j}
                                              \big)\text{,}
\end{equation}
with $u_{pair}$ being defined in eqn~\eqref{eqn:pair}.
Since the shape of each particle is modeled as a disk, we set the translational diffusion tensor $\mathbb{D} = D_t \mathbb{I}$, where $D_t$ is the (isotropic) translational diffusion constant and $\mathbb{I}$ is the $2 \times 2$ identity matrix.
Correspondingly, $D_r$ denotes the rotational diffusion constant. 
As described in ref. \citen{Liao2018}, we find that the relationship between these two diffusion constants of a hard sphere in the low Reynolds number regime is given \text{via} $D_r = 3 D_t / \sigma_h^2$.
We also define the diameter of an effective (eff) hard sphere \text{via}
$\sigma_{eff} = 
    \int_{0}^{\infty}
      \left(
        1 -
        \text{exp}
          \left[
            - \beta u_{sr}(r)
          \right]
      \right)
      \text{d}r$.
With the repulsive strength $\epsilon^* = 10$ considered in the present system, $\sigma_{eff} \approx 1.07851 \sigma$.
By choosing $\sigma_h = \sigma_{eff}$, we obtain $D_r = 2.57914 D_t / \sigma^2$.
%

%
In eqn~\eqref{eqn:coupled_Langevin_trans}, the effective propulsion force, which drives the active motion, is given by $\boldsymbol{F}_i = F_0 \hat{\boldsymbol{e}}_i$ with $\hat{\boldsymbol{e}}_i = \hat{\boldsymbol{\mu}}_i$. 
For simplicity, in the remainder of this work we present the impact of the effective propulsion force via the motility $v_0 = \beta D_t F_0$.
Finally, the random force $\boldsymbol{\xi}_{i}\left(t\right)$ and torque $\Gamma_{i}\left(t\right)$ for the $i$th particle are zero-mean Gaussian white noise, which satisfy 
$\langle
   \boldsymbol{\xi}_{i}(t)
   \otimes
   \boldsymbol{\xi}_{j}(t')
 \rangle =
   2  \delta_{ij}  
      \delta(t-t')
      \mathbb{I} /
        (D_t\beta^{2})$ 
and 
$\langle
  \Gamma_i(t)
  \Gamma_j(t')
\rangle = 
  2  \delta_{ij}
     \delta(t-t') /
       (D_r\beta^{2})$. 
The angle brackets $\langle \cdots \rangle$ denote ensemble average, and the symbol $\otimes$ represents dyadic product. 
%

%
Equations~\eqref{eqn:coupled_Langevin_trans} and~\eqref{eqn:coupled_Langevin_rot} are solved via the Euler-Maruyama method \cite{Kloeden1992} with the discrete time step $\Delta t = 2 \times 10^{-5} \tau$, where $\tau$ is the Brownian diffusion time, given by $\tau = \sigma^2/D_t$.
We choose a quadratic box with side length $L_x = L_y = L$, and we use periodic boundary conditions in both directions. 
In order to treat the long-range dipole-dipole interactions, we employ a two-dimensional (2D) Ewald summation, as outlined in Appendix~\ref{sec:Ewald}.
The particle number is set to $N = 1156$, unless otherwise stated.
All simulations are started with randomly oriented particles located on a square lattice.
A typical run then consists of at least $5 \times 10^5$ steps for reaching a steady state, followed by a production period of $5 \times 10^5$ steps.
The statistical properties of the dipolar particles (see Sec.~\ref{sec:target_quant}) are measured every $500$ steps.
The simulations are carried out at three values of the mean area fraction $\Phi = N \pi \sigma_{eff}^2 /\left(4L^2\right)$, where the area of a single particle is defined as $\pi \sigma_{eff}^2 / 4$.
Specifically, we consider the values $\Phi = 0.12$, $\Phi = 0.23$, and $\Phi = 0.58$ (for the exact values, see note \citen{phi_alternative}).
To investigate the impact of activity, we perform simulations at different dimensionless propulsion speeds $v_0^* = v_0 \sigma/ D_t$ and various dipolar coupling strengths $\lambda$.
We note that $v_0^*$ is indeed the same as the commonly used P\'eclet number $Pe$. \cite{Buttinoni2013, Cates2015, Blaschke2016}

\subsection{\label{sec:target_quant}Target quantities}
In this section we introduce the target quantities which will later be used to characterize the system's behavior.
The choice of these quantities is inspired by earlier research on non-dipolar active systems on the one hand, and passive dipolar systems on the other hand.
\subsubsection{\label{cluster}Clustering behavior}
%
%
%
%
%
%
%
It is well established that active particles with purely repulsive interactions have a tendency to form large clusters and even phase-separate if the motility is sufficiently high (motility-induced phase separation). \cite{Tailleur2008, Bialke2013, Cates2015, Digregorio2018}
We would therefore expect that the present system displays similar behavior at least at low dipolar coupling strengths, \textit{i.e.}, small values of $\lambda$.
To characterize the clustering behavior of the dipolar active particles, we perform a cluster size analysis based on a simple distance criterion: 
Two particles are regarded as being in contact if their center-to-center distance is smaller than a distance $r_{L}$. 
A cluster is then a set of particles that are in contact with each other. 
We quantify the cluster formation by 
\begin{equation}
  \phi_c = \left\langle n_{lcl} \right\rangle / N \text{, }
  \label{eqn:phi_c}
\end{equation}
where $n_{lcl}$ denotes the number of particles in the largest cluster.
The quantity $\phi_c$ approaches $0$ in a state with particles being ``essentially uncorrelated'', where not only the one-particle density is a constant, but also the particle correlations (measured by $g(r)$) are weak. 
Further, $\phi_c$ is close to zero in a state with finite-sized chains or clusters with a size much smaller than $N$.
In contrast, $\phi_c \approx 1$ when the active particles form ``giant'' clusters with a size comparable to $N$. \cite{Buttinoni2013}
We choose $r_{L}$ as the distance corresponding to the first peak in the radial distribution function of a corresponding ``reference system'' ($v_0^* = 0$, $\lambda = 0$) at the given density. 
In this way we can systematically investigate the impact of the motility and dipolar coupling without changing the cluster criteria. 
As a result, we obtain $r_{L} \approx 1.16 \sigma$ for $\Phi = 0.12$, $r_{L} \approx 1.13 \sigma$ for $\Phi = 0.23$, and $r_{L} \approx 1.12 \sigma$ for $\Phi = 0.58$ (for the exact values, see note \citen{rL}).

\subsubsection{\label{sec:order}Global orientational ordering}
%
The appearance of global orientational order in passive systems of dipolar particles has a long history.
In three dimensions, the occurrence of ferromagnetic liquid and solid states has been confirmed both by computer simulations \cite{Weis2005} and by theory. \cite{Weis2005a}
In spatially confined systems, the situation is less clear.
A ferromagnetic state has indeed been observed in slab-like systems composed of three and more layers of dipolar particles. \cite{Klapp2002, Trasca2008}
Systems with less than three layers have been considered in ref. \citen{Trasca2008}.
There, one did not find clear hints for the ferromagnetic order in very thin films, consistent with other studies. \cite{Weis2002, Weis2003}
%

%
The behavior of the global orientational ordering has also been investigated in models of self-propelled particles with ``velocity-alignment'' interactions, such as the Vicsek model \cite{Vicsek1995} and its variants (see, \textit{e.g.}, ref. \citen{Martin-Gomez2018}). 
In these systems, one observes so-called flocking states, where particles gather together and move collectively toward a certain direction. 
In the present system, the orientational (dipolar) interaction is more complicated (than the Heisenberg-like interactions in the Vicsek model); still, one could imagine that activity-induced flocking states also emerge in the dipolar active system. 
To this end we consider (as is common in the literature on flocking states \cite{Martin-Gomez2018}) the parameter
\begin{equation}
  \phi_{\boldsymbol{e}}
  = 
  \dfrac{1}{N} 
  \left\langle 
    \left\vert 
      \sum_{i=1}^{N}
      \hat{\boldsymbol{e}}_i
    \right\vert
  \right\rangle \text{,} \label{eqn:phi_e}
\end{equation}
which corresponds to the magnitude of the average orientation. 
This order parameter is unity if all particles self-propel toward the same direction, and zero if the particle orientations are uncorrelated. 

\subsubsection{\label{sec:chain}Chain formation}
%
At low densities, passive particles with strong dipole-dipole interactions ($\lambda > 1$) are known to self-assemble into chains and rings. \cite{Weis1993, Duncan2004, Weis2005, Duncan2006, Rovigatti2011, Rovigatti2012, Rovigatti2013, Kantorovich2013, Kantorovich2016, Ronti2017, Camp2018}
The reason is easily seen from eqn~\eqref{eqn:Udd}:
Considering two dipolar spheres separated by a distance $\sigma$, the configuration with the lowest energy is the head-to-tail configuration (with $u_{dd} = - 2 \mu^2\sigma^{-3}$). 
In contrast, if the particles are arranged side by side, the energy reaches a maximum for parallel orientation (with $u_{dd} = \mu^2\sigma^{-3}$), suggesting that such a configuration is energetically unfavorable. 
%

%
To characterize chain formation in the dipolar active system, we use different strategies depending on the density regime considered.
The first strategy is adequate for low densities ($\Phi \lesssim 0.23$), where the chains can essentially be considered as isolated objects.
In this situation, we consider a chain as a set of, at least, three particles which are mutually ``bonded.''
Specifically, two particles $i, j$ are regarded as ``bonded'' in the chain, if the following criteria are fulfilled:
$\vert \boldsymbol{r}_{ij} \vert \leq r_{p}$, $\hat{\boldsymbol{\mu}}_{i} \cdot \hat{\boldsymbol{\mu}}_{j} > 0 $, and $\big( \hat{\boldsymbol{\mu}}_{i} \cdot \boldsymbol{r}_{ij} \big)\big( \hat{\boldsymbol{\mu}}_{j} \cdot \boldsymbol{r}_{ij} \big) > 0$.
Here, $r_{p} = 1.25 \sigma$ is set to a distance between the location of the first peak and the first valley of the pair correlation function. \cite{Peroukidis2016, Schmidle2012, Schmidle2013}
%
%
Based on these rules, we quantify the low-density chain formation via the parameter
\begin{equation}
  \phi_{p} = \left\langle N_{p} \right\rangle / N \text{, } \label{eqn:phi_p}
\end{equation}
where $N_{p}$ is the total number of particles which reside in chains.
In the context of aggregating molecular systems, $\phi_p$ is often called the degree of polymerization.
%

%
At high densities (\textit{e.g.}, $\Phi = 0.58$), the particles are obviously closer to each other, and the string-like structures are no longer isolated.
This causes the degree of polymerization $\phi_{p}$ to no longer be appropriate to characterize the self-assembly.
To overcome this difficulty, we propose a different order parameter to describe the chain formation quantitatively. 
The starting point is the dipole-dipole correlation function
\begin{equation} 
 g_{_{\boldsymbol{\mu}}}(r_{\bot},r_{\|}) = 
 \dfrac{
   \left\langle 
     \sum_{i \neq j}
     \hat{\boldsymbol{\mu}}_i \cdot \hat{\boldsymbol{\mu}}_j 
     \delta\big( r_{\bot} - r_{ij}^{\bot} \big)
     \delta\big( r_{\|}   - r_{ij}^{\|}   \big) 
   \right\rangle
 }
 {
   \left\langle
     \sum_{i \neq j}
     \delta\big( r_{\bot} - r_{ij}^{\bot} \big)
     \delta\big( r_{\|}   - r_{ij}^{\|}   \big) 
   \right\rangle
 } \text{, } \label{eqn:g_mu_xy}
\end{equation}
where the transverse and longitudinal displacement of particle $j$ relative to $i$ is $r_{ij}^{\|} = \boldsymbol{r}_{ij} \cdot \hat{\boldsymbol{\mu}}_i$ and $r_{ij}^{\bot} = \sqrt{ \vert \boldsymbol{r}_{ij} \vert^{2} - \left(\boldsymbol{r}_{ij} \cdot \hat{\boldsymbol{\mu}}_i\right)^{2}}$, respectively. 
To extract the angular dependence at short distances, we transform the Cartesian coordinates employed in eqn~\eqref{eqn:g_mu_xy} to polar coordinates by using $r = \sqrt{r_{\bot}^{2} + r_{\|}^{2}}$ and $\theta = \text{atan2}\big( - r_{\bot}, \: r_{\|} \big)$, and compute the function
\begin{equation}
  \tilde{g}_{_{\boldsymbol{\mu}}}\left(\theta\right) = 
  \dfrac{ 
    \int_{0}^{r_s}g_{_{\boldsymbol{\mu}}}(r,\theta)r\text{d}r
  }
  {
    \int_{0}^{r_s}r\text{d}r
  } \text{. }
  \label{eqn:g_mu_theta}
\end{equation}
Here, the distance $r_s$ is set to $r_s = L / 8 \approx 5 \sigma$ such that $r_s$ is much smaller than the half of the box size (which corresponds to the length scale of the periodic boundary conditions).
In the presence of chain-like structures with head-to-tail alignment of neighboring particles, we expect the angular correlation function, $\tilde{g}_{\boldsymbol{\mu}}\left(\theta\right)$, to display positive maxima both, in front of and behind the reference particle (\textit{i.e.}, at $\theta = 0$ or $\pi$). 
In contrast, the dipole moments of the particles on the right- and left-hand side of the reference particle should remain rather uncorrelated (\textit{i.e.}, $\tilde{g}_{_{\boldsymbol{\mu}}}\left(\theta\right) \approx 0$ for $\theta = \pi/2$ or $- \pi/2$).
With this picture in mind, we measure the quantity
\begin{equation}
  \mathzapf{Z} = 
    \tilde{g}_{_{\boldsymbol{\mu}}}\left(\theta_{max}\right)
    -
    \tilde{g}_{_{\boldsymbol{\mu}}}\left(\theta_{min}\right)
  \text{, }
  \label{eqn:Z}
\end{equation}
where $\tilde{g}_{_{\boldsymbol{\mu}}}\left(\theta\right)$ reaches its maximum and minimum, respectively, at $\theta = \theta_{max}$ and $\theta = \theta_{min}$.
By properly choosing a threshold value $\mathzapf{Z}_{thres}$, we define the dipolar particles as exhibiting chain-like structure when $\mathzapf{Z} \geq \mathzapf{Z}_{thres}$. 
As will be discussed in more detail in Sec.~\ref{sec:chain}, a reasonable value for $\mathzapf{Z}_{thres}$ at $\Phi = 0.58$ is $\mathzapf{Z}_{thres} = 0.17$.
\begin{table}[!htbp]
    \small
    \setlength{\tabcolsep}{5pt} 
    \begin{tabular}{| p{2.1cm} | c | c | c | c |}
    \hline
    State & Clustering & Orientational & \multicolumn{2}{c|}{Chain formation} \\ \cline{4-5}
          &            &               &                  &                   \\[-0.25cm] 
          &            & ordering      & \parbox[b]{1.1cm}{$\Phi \lesssim 0.23$} & $\Phi = 0.58$ \\ \hline \hline
    Homogeneous, isotropic fluids      & $\phi_c \leq 0.5$ & $\phi_{\boldsymbol{e}} \leq 0.5$ &  $\phi_p \leq 0.5$ & $\mathzapf{Z} \leq 0.17$ \\ \hline
    Chain-like structures       & $\phi_c \leq 0.5$ & $\phi_{\boldsymbol{e}} \leq 0.5$ &  $\phi_p > 0.5$ & $\mathzapf{Z} >    0.17$ \\ \hline
    Micro-flocking              & $\phi_c \leq 0.5$ & $\phi_{\boldsymbol{e}} >    0.5$ & \multicolumn{2}{c|}{ - }\\ \hline
    Macro-flocking              & $\phi_c >    0.5$ & $\phi_{\boldsymbol{e}} >    0.5$ & \multicolumn{2}{c|}{ - }\\ \hline
    Motility-induced clustering & $\phi_c >    0.5$ & $\phi_{\boldsymbol{e}} \leq 0.5$ & \multicolumn{2}{c|}{ - }\\
    \hline
    \end{tabular}
    \caption[Characterization of the states of dipolar active particles]{Characterization of the states of dipolar active particles according to the order parameters defined in Sec.~\ref{sec:target_quant}.}
    \label{table:states}
\end{table}
%

%
\section{\label{sec:result}Simulation results}
%
%
Based on the target quantities described in Sec.~\ref{sec:target_quant}, we can classify the states observed in our BD simulations performed at three densities and at various values of the motility $v_0^*$ and the dipolar coupling strength $\lambda$.
In the following Sec.~\ref{sec:chaining} $-$~\ref{sec:clustering} we discuss, for each density, the state diagram in the ($v_0^*$, $\lambda$) plane.
Specifically, we identify five states whose characteristics are summarized in Table~\ref{table:states}.
\begin{figure}[!htbp]
  \centering
  \includegraphics[width=0.9\linewidth]{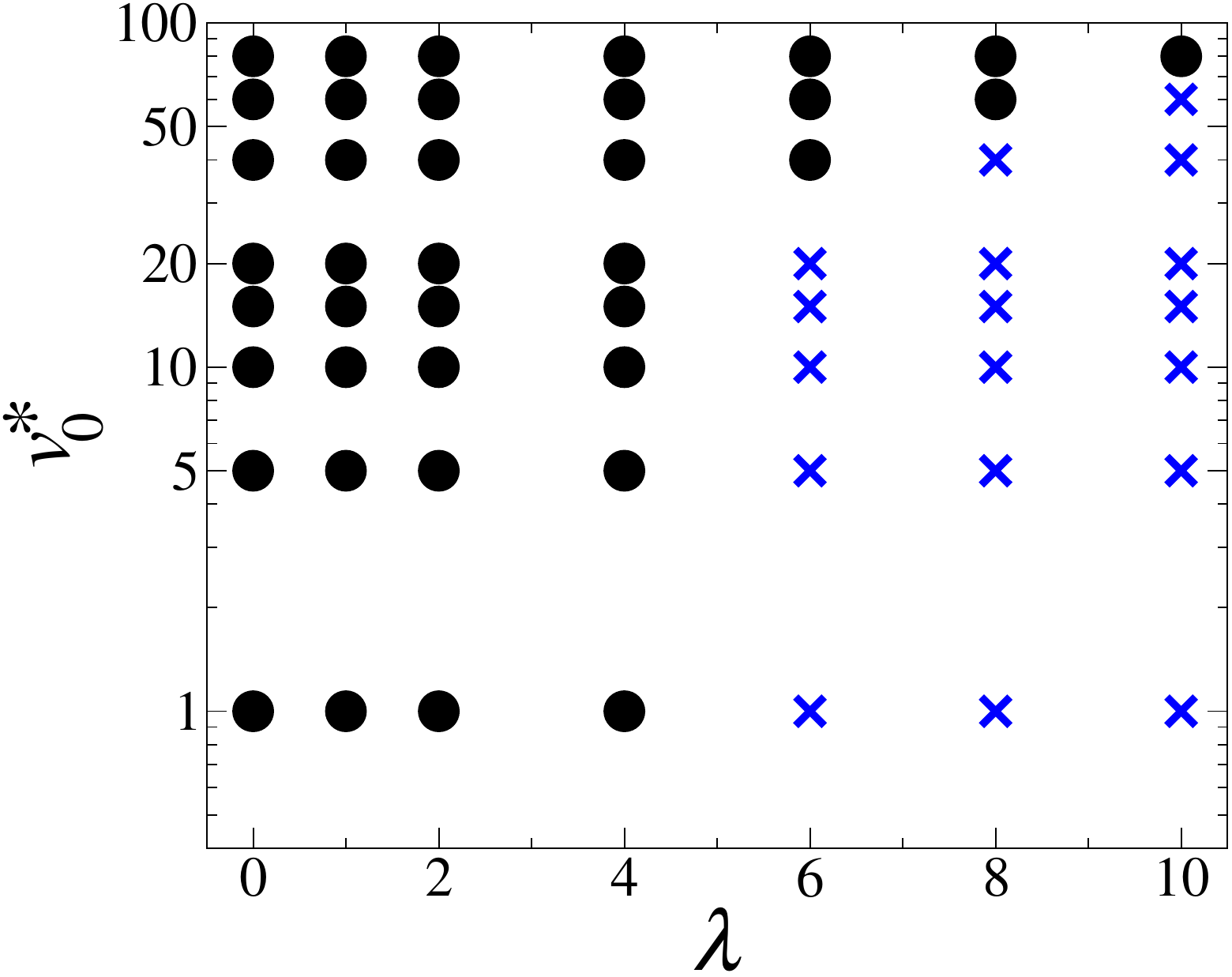}
  \caption[The state Diagram at $\Phi = 0.12$]{(Color online) State diagram of dipolar active particles in the ($v_0^*$, $\lambda$) plane at $\Phi = 0.12$. The points on the diagram indicate the parameter combinations used in the simulations. At $\Phi = 0.12$, we have observed homogeneous, isotropic fluid states (black circles) and chain-like structures (blue crosses). }
  \label{fig:P01StateDiagram}
\end{figure}

\subsection{\label{sec:chaining}The low density regime ($\Phi = 0.12$)}

\begin{figure}[!htbp]
  \centering
  \includegraphics[width=1.0\linewidth]{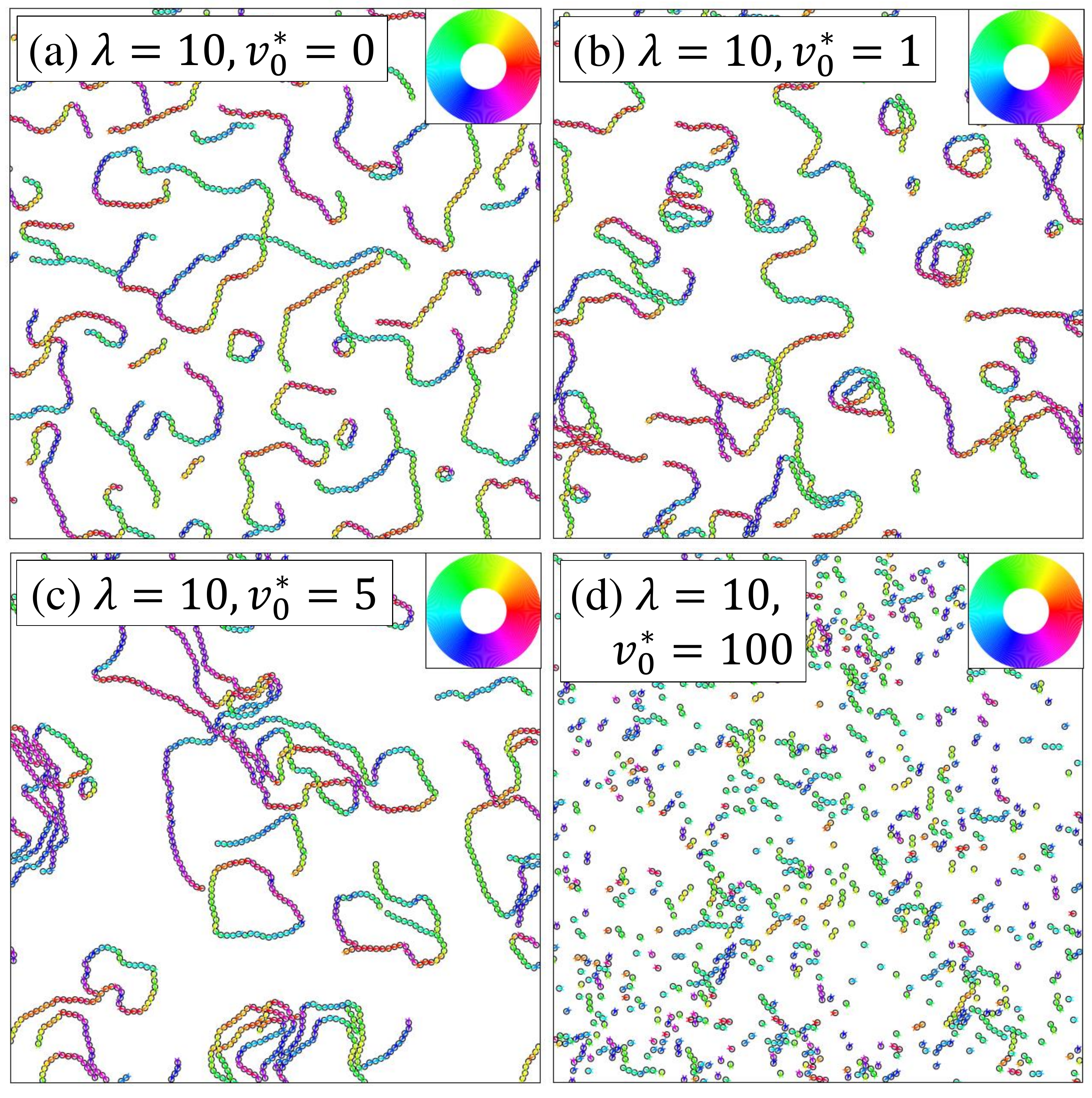}
  \caption[Snapshots at $\Phi = 0.12$]{(Color online) Representative simulation snapshots at $\Phi = 0.12$. Particles are colored according to their dipole orientations as indicated by the color ring in the inset. }
  \label{fig:P01Snapshots}
\end{figure}

We start by considering the state diagram in the low-density regime, taking $\Phi = 0.12$ as a representative example, see Fig.~\ref{fig:P01StateDiagram}.
In the absence of dipolar coupling ($\lambda = 0$), non-dipolar active particles at such a low density typically display a homogeneous isotropic fluid state, without significant translational correlations or orientational ordering. \cite{Buttinoni2013}
As long as the dipolar coupling strength remains relatively small ($\lambda \lesssim 4$), the system still shows the same homogeneous isotropic fluid state for all motilities considered.
In contrast, when $\lambda$ exceeds a value of about $6$, the dipolar active particles form chain-like structures provided that the motility is below a critical motility $v_{0, c}^*\left(\lambda\right)$, whose value depends on $\lambda$.
Above this critical motility $v_{0, c}^*\left(\lambda\right)$, the chain-like structures found at smaller motilities break, and the system becomes homogeneous and isotropic.
As Fig.~\ref{fig:P01StateDiagram} reveals, $v_{0, c}^*\left(\lambda\right)$ increases with increasing $\lambda$.
As a visualization of the impact of motility on dipolar active particles at a high coupling strength, such as $\lambda = 10$, we plot in Fig.~\ref{fig:P01Snapshots} representative snapshots.
At $v_0^* = 0$, nearly all particles are bound into chains and rings, as it is typical for passive dipolar particles in 2D. \cite{Tavares2002}
With increasing $v_0^*$ the ring structures are seemingly more abundant, as can be seen in Fig.~\ref{fig:P01Snapshots}(a-c).
Further increase in $v_0^*$ eventually causes the chain-like structures to break into fragments of short chains and single particles, as shown in Fig.~\ref{fig:P01Snapshots}(d).

\begin{figure}[!htbp]
  \centering
  \includegraphics[width=0.9\linewidth]{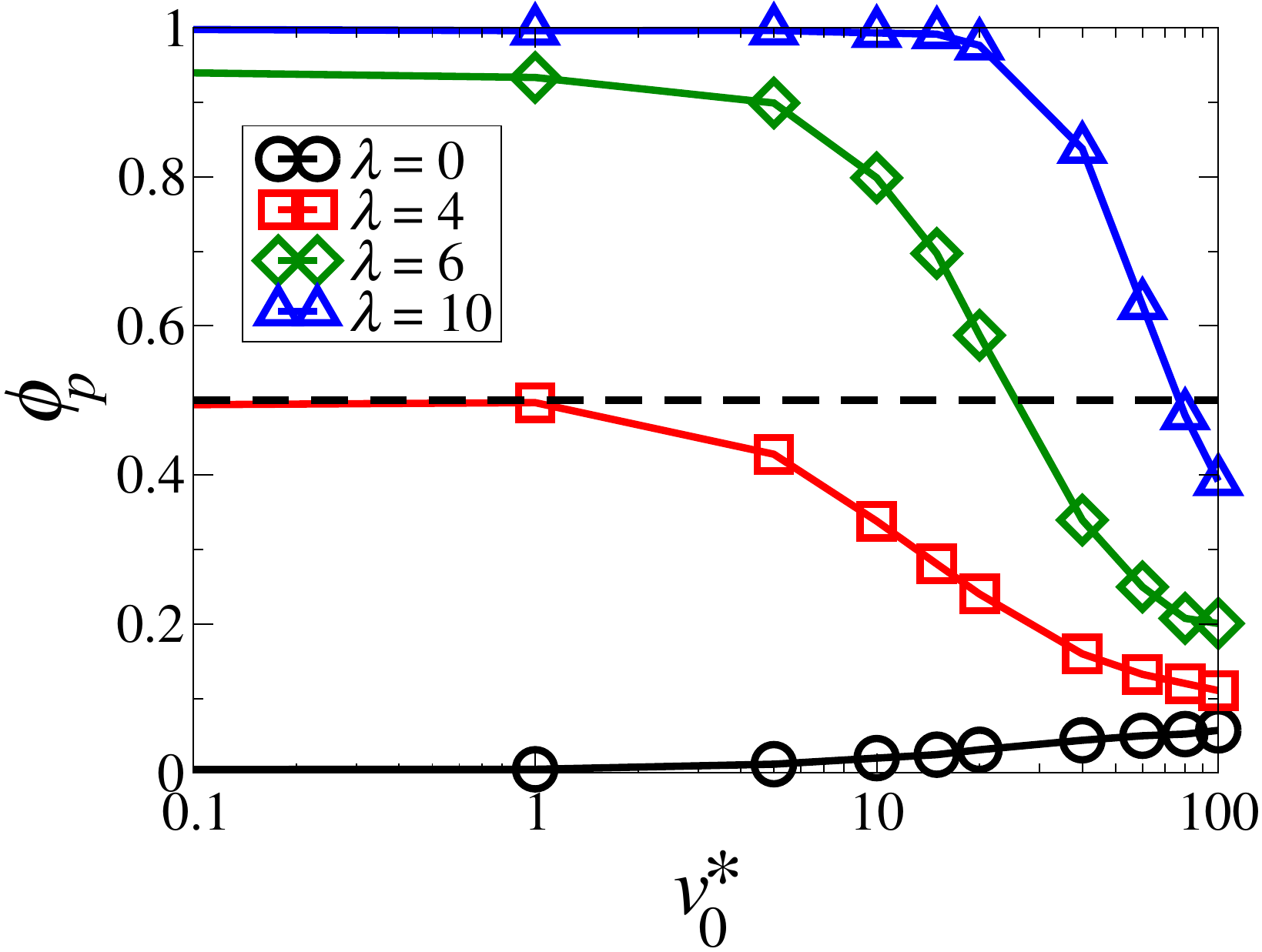}
  \caption[The order parameter $\phi_p$ at $\Phi = 0.12$]{(Color online) Degree of polymerization $\phi_p$ as a function of the motility $v_0^*$ at $\Phi = 0.12$ for the coupling strength $\lambda = 0$ (black dots), $4$ (red squares), $6$ (green diamonds), and $10$ (blue triangles). The dashed horizontal line at $\phi_p = 0.5$ marks our criterion for a state with chain-like structures (see Table~\ref{table:states}). The solid lines are guides to the eye.}
  \label{fig:phi_p}
\end{figure}

%
To quantitatively characterize the motility-induced destruction of chain-like structures, we plot in Fig.~\ref{fig:phi_p} the degree of polymerization $\phi_p$ as a function of the motility $v_0^*$ for various coupling strengths $\lambda$.
In the absence of dipole-dipole interactions ($\lambda = 0$), particles do not self-assembly into chain-like structures, still we observe a slight increase in $\phi_p$ [see eqn~\eqref{eqn:phi_p}] upon increasing motilities $v_0^*$.
This can be attributed to the dynamical clustering of active Brownian particles into small (finite-sized) aggregates. \cite{Liao2018, Theurkauff2012, Buttinoni2013}
Indeed, if the particles inside these finite-sized clusters fulfill our criteria of being bonded, the order parameter $\phi_p$ will be non-zero (yet small), despite the fact that dipole-dipole interactions are absent.
Once the dipole-dipole interactions are introduced, pronounced chain-like structures appear for coupling strengths $\lambda \gtrsim 6$ and low motilities $v_0^*$, giving rise to large values of $\phi_p$ ($\phi_p \approx 1$).
Upon an increase in the motility, the degree of polymerization $\phi_p$ gradually decreases, reflecting that the self-propulsion opposes the formation of chain-like structures.
This behavior resembles that seen in passive, dilute systems of self-assembled dipolar particles, when the temperature is increased. \cite{Kantorovich2015}
In this sense, the particle motility in the active dipolar system may be viewed as an analog to the temperature in the passive system.
%

\begin{figure}[!htbp]
  \centering
  \includegraphics[width=0.9\linewidth]{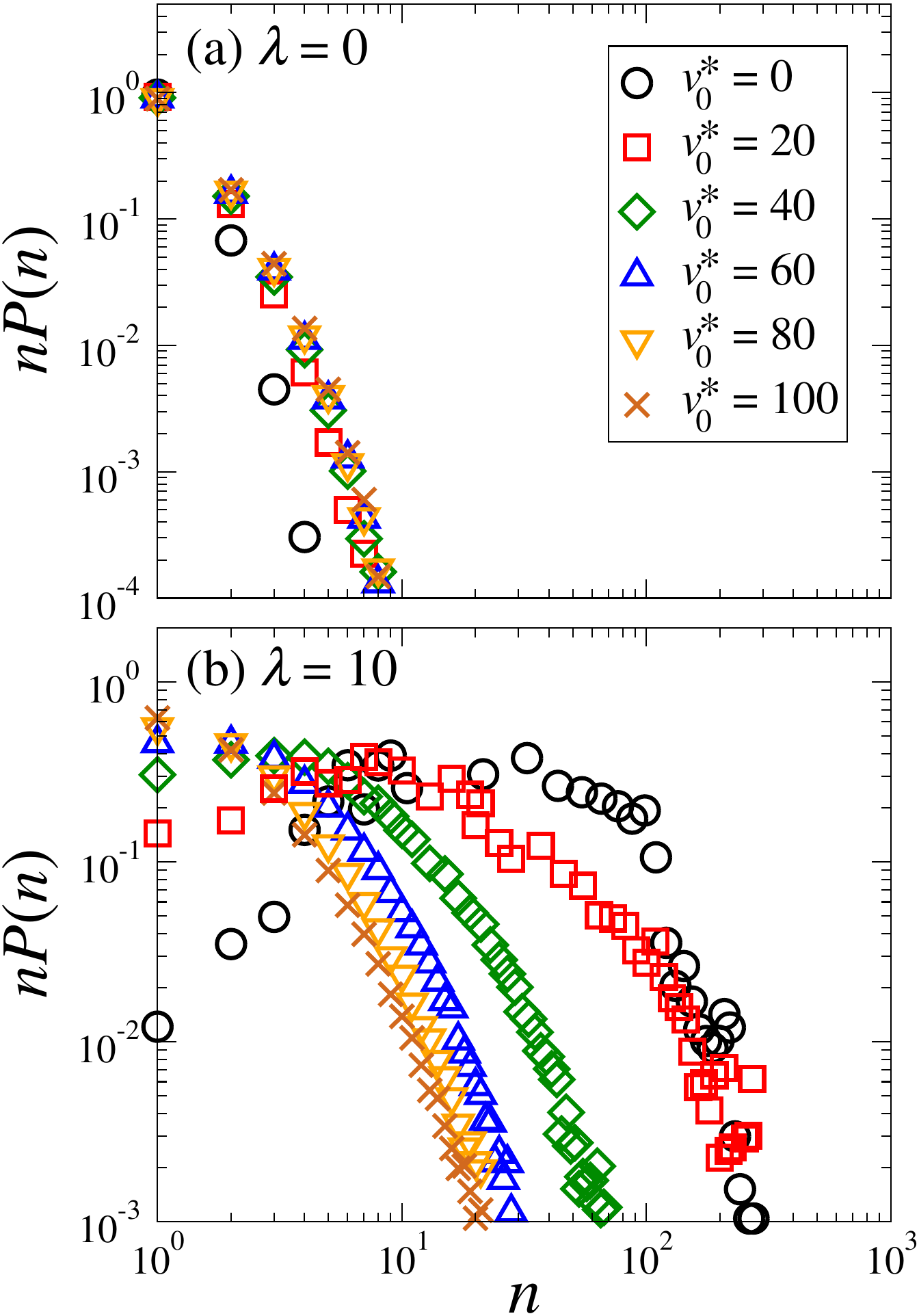}
  \caption[The weighted distribution function of chain size at $\Phi = 0.12$]{(Color online) Weighted distribution of the chain size at $\Phi = 0.12$ for motilities $v_0^* = 0$ (black circles), $20$ (red squares), $40$ (green diamonds), $60$ (blue triangles up), $80$ (orange triangles down), and $100$ (brown crosses) with the coupling strength $\lambda = 0$ (a) and $\lambda = 10$ (b).}
  \label{fig:chainDistri}
\end{figure}

%
The degree of polymerization $\phi_p$ only describes the chaining behavior globally. 
To supplement our analysis of chain formation, we measure the chain size distribution function $P(n)$, which represents the probability that a randomly selected chain consists of $n$ particles.
In the presence of chain formation, the simulated systems usually contain several long chains and many short chains, such that $P(n)$ becomes very small at large $n$.
Therefore, we plot in Fig.~\ref{fig:chainDistri} the chain size distribution weighted by $n$, where $nP(n)$ is proportional to the probability that a randomly selected particle belongs to a chain with a size $n$.
As seen in Fig.~\ref{fig:chainDistri}(a), the weighted distribution curves of non-dipolar particles vanish at around a chain size $n \lesssim 10$, and the values slightly increase with increasing motilities.
This is consistent with the aforementioned dynamical clustering.
The corresponding functions at large dipolar coupling strengths ($\lambda = 10$) look very different, as shown in Fig.~\ref{fig:chainDistri}(b).
Here, we observe a broad peak between $10 \lesssim n \lesssim 100$ at $v_0^* = 0$, reflecting formation of long chains.
Upon an increase in the motility, this peak is gradually shifted to a smaller chain size. 
Finally, this peak disappears once $v_0^* \gtrsim 60$. 
This suggests a vanishing of chains with the most probable size at large values of $v_0^*$.
%

%
Inspecting again Fig.~\ref{fig:P01Snapshots}(d) we see that even at the largest coupling strength considered, the dipole moments do not align on a length scale comparable to that of the simulation box, regardless of the values of $v_0^*$.
In other words, there is no pronounced global orientational order.
Indeed, through measuring the orientational order parameter $\phi_{\boldsymbol{e}}$, we confirm that $\phi_{\boldsymbol{e}}$ remains small ($\phi_{\boldsymbol{e}} < 0.5$) for the parameter combinations explored in Fig.~\ref{fig:P01StateDiagram}, indicating that the systems are globally isotropic at low densities.
\begin{figure}[!htbp]
  \centering
  \includegraphics[width=0.85\linewidth]{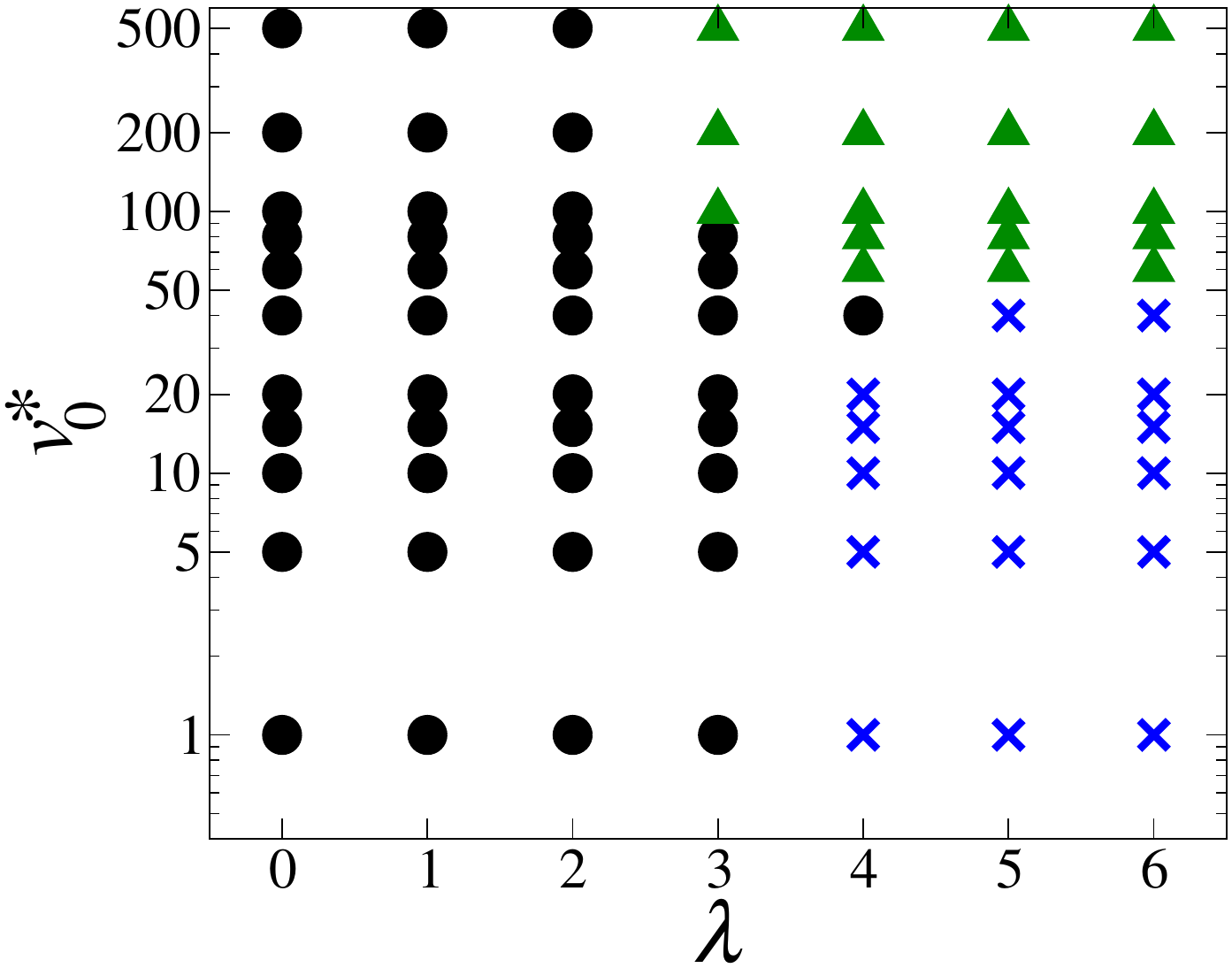}
  \caption[The state Diagram at $\Phi = 0.23$]{(Color online) State diagram of dipolar active particles in the ($v_0^*$, $\lambda$) plane at $\Phi = 0.23$. The points on the diagram indicate the parameter combinations used in the simulations. At $\Phi = 0.23$, we have observed homogeneous, isotropic fluid states (black circles), micro-flocking (green triangles), and chain-like structures (blue crosses). }
  \label{fig:P02StateDiagram}
\end{figure}
\begin{figure}[!htbp]
  \centering
  \includegraphics[width=1.0\linewidth]{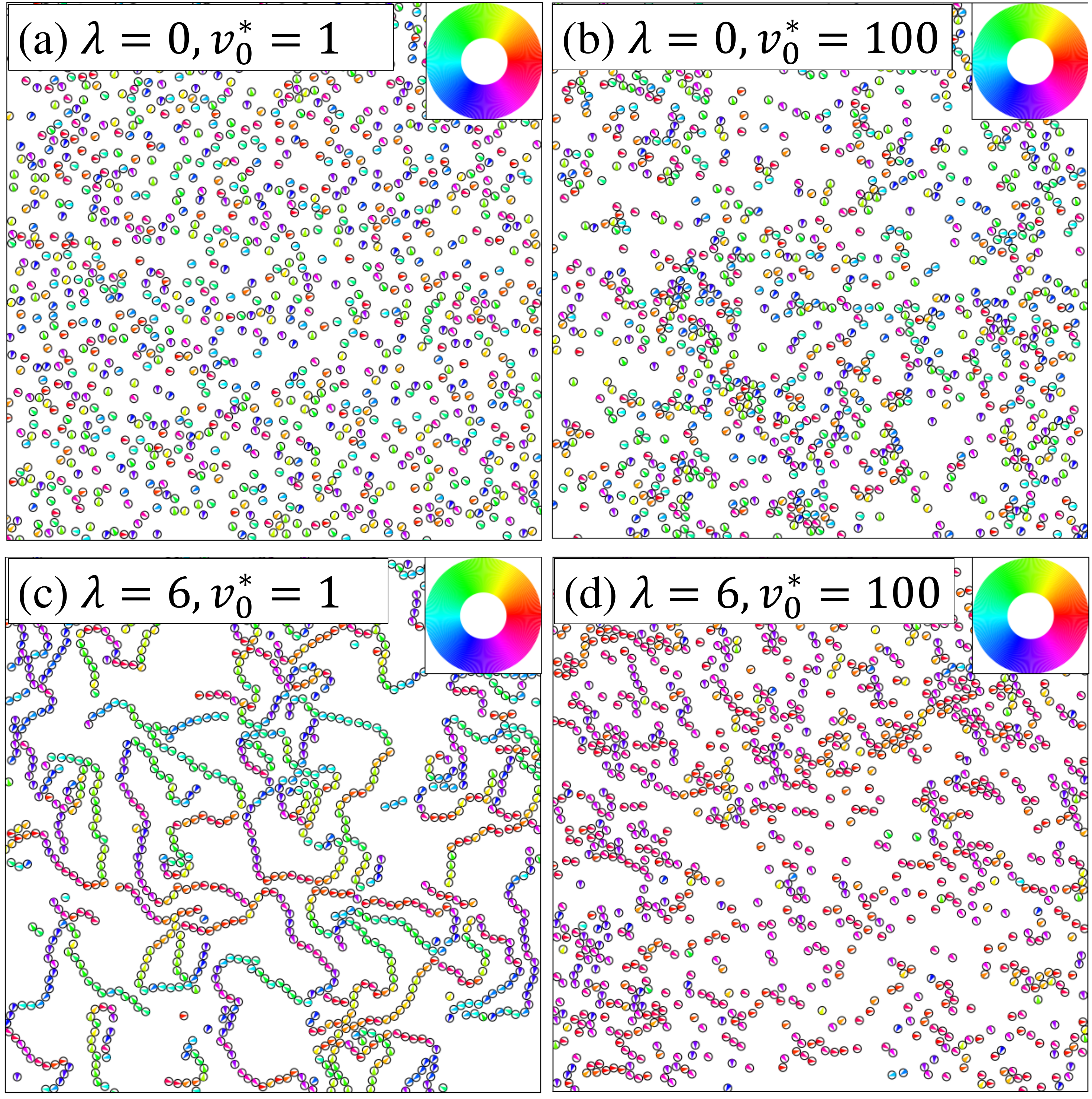}
  \caption[Snapshots at $\Phi = 0.23$]{(Color online) Representative simulation snapshots at $\Phi = 0.23$. Particles are colored according to their orientations as indicated by the color ring in the inset. }
  \label{fig:P02Snapshots}
\end{figure}

\subsection{\label{sec:ordering}The intermediate density regime ($\Phi = 0.23$)}

%
At the density $\Phi = 0.23$, the state diagram (see Fig.~\ref{fig:P02StateDiagram}) at small and intermediate motilities ($v_0^* \lesssim 20$) is very similar to that at $\Phi = 0.12$ (see Fig.~\ref{fig:P01StateDiagram}):
For weakly coupled systems ($\lambda \lesssim 3$), we observe a homogeneous and isotropic fluid state, whereas strong dipolar coupling leads to chain-like structures.
This is visualized by the simulation snapshots presented in Fig.~\ref{fig:P02Snapshots}(a) and (c).
For these ``nearly passive'' systems, the main difference compared to the dilute systems in Fig.~\ref{fig:P01StateDiagram} is that the change from the homogeneous isotropic fluid state into the chain-like state occurs at a somewhat smaller value of $\lambda$.
This is plausible, because the higher density leads to a smaller mean separation between the particles and thus, to a higher probability for chain formation.
%

%
We now turn to the impact of the motility.
At small values of $\lambda$, an increase in $v_0^*$ toward high values leads to the formation of finite-sized clusters [see Fig.~\ref{fig:P02Snapshots}(b)] which are, however, not stable.
Thus, the system remains homogeneous and isotropic on average.
This finding is consistent with earlier research on active Brownian particles.
In particular, the phenomenon of motility-induced phase separation, accompanied by formation of ``giant'' clusters ($\phi_c \approx 1$), is known to occur only at higher densities. \cite{Buttinoni2013}
As Fig.~\ref{fig:P02StateDiagram} reveals, the picture changes if the dipolar coupling becomes larger.
Consider, \textit{e.g.}, the case $\lambda = 6$.
Increasing the motility from zero, the chain-like structures observed at small $v_0^*$ break [see Fig.~\ref{fig:P02Snapshots}(c)], and the particles instead form small clusters characterized by the same orientation [see Fig.~\ref{fig:P02Snapshots}(d)].
More quantitatively, the clustering parameter $\phi_c$ remains small ($\phi_c < 0.5$) but the orientational order parameter $\phi_{\boldsymbol{e}}$ reaches values above $0.5$.

\begin{figure}[!htbp]
  \centering
  \includegraphics[width=0.9\linewidth]{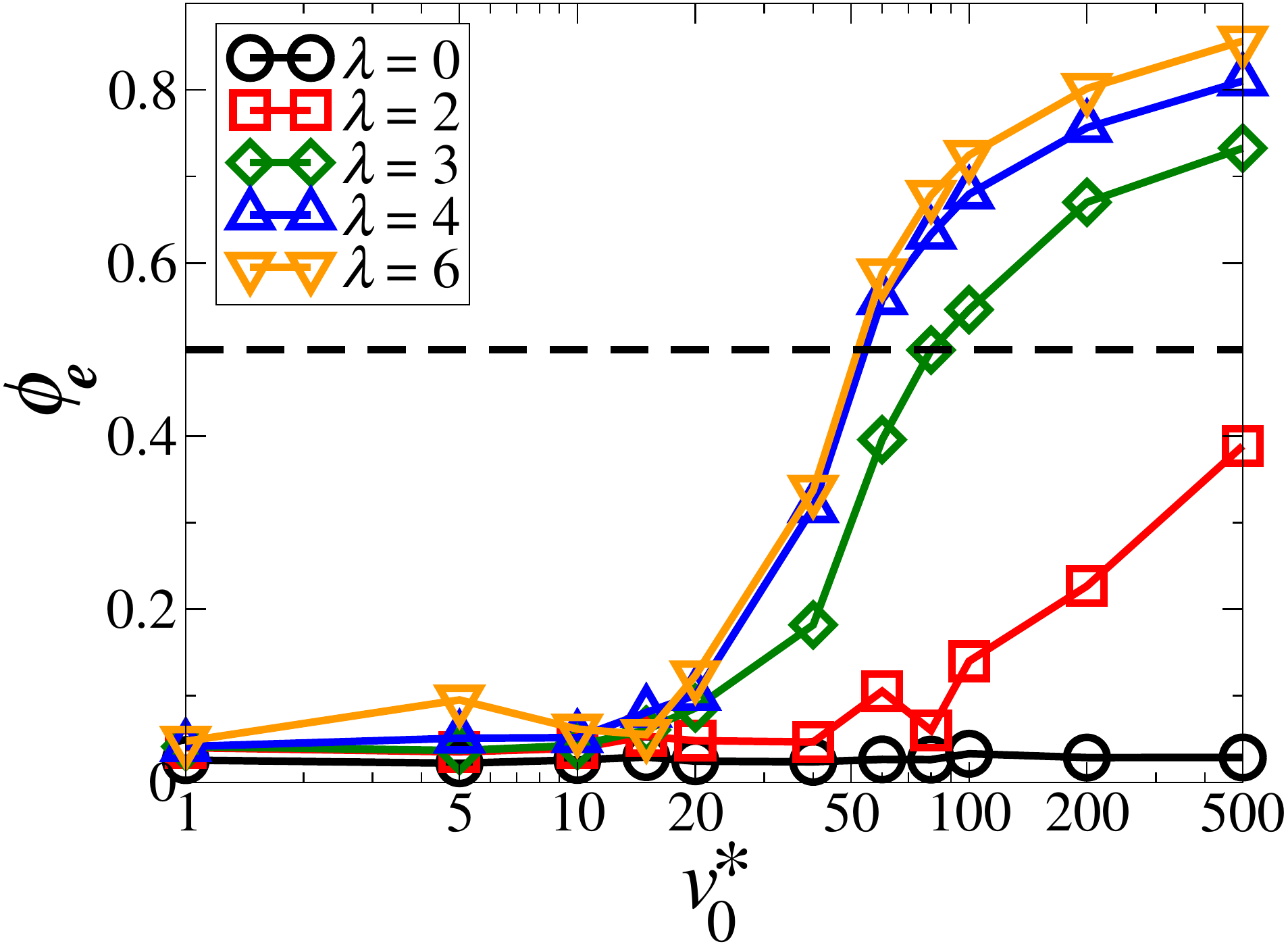}
\caption[The order parameter $\phi_{\boldsymbol{e}}$ at $\Phi = 0.23$]{(Color online) Global polarization $\phi_{\boldsymbol{e}}$ (open symbols) as a function of the motility $v_0^*$ at $\Phi = 0.23$ for the coupling strength $\lambda = 0$ (black circles), $2$ (red squares), $3$ (green diamonds), $4$ (blue triangles up), and $6$ (orange triangles down). The dashed line indicates the value corresponding to the emergence of flocking states (see Table~\ref{table:states}). The solid lines are guides to the eye.}
\label{fig:P02phi_e}
\end{figure}

%
The full behavior of the orientational order parameter $\phi_{\boldsymbol{e}}$ as function of $v_0^*$ is shown in Fig.~\ref{fig:P02phi_e}.
For strongly coupled systems ($\lambda \geq 3$), the order parameter abruptly changes from (essentially) zero to large values at a ``critical'' motility $v_{0, c}^*\left(\lambda\right) \approx 20$.
We take this as an indication for a motility-induced formation of ordered, yet small, clusters.
According to Table~\ref{table:states}, this is characteristic of a ``micro-flocking'' state.
We recall that, for active particles, the development of average alignment between neighboring particles implies that they move along the same direction.
To explore whether these small clusters are indeed ``microscopic'' structures, we measured the cluster size distribution.
It turns out that the characteristic cluster size $n_0$ does not scale up with the system size $N$ (see the detailed analysis in Appendix~\ref{sec:clusterDistri}).
%

%
Inspecting Fig.~\ref{fig:P02phi_e} again, it is found that for systems at an intermediate coupling strength $\lambda = 2$, the order parameter $\phi_{\boldsymbol{e}}$ gradually increases from zero as $v_0^* \gtrsim 100$, suggesting that the ``micro-flocking'' state might still appear. 
However, further increase in $v_0^*$ requires to employ even smaller simulation time steps $\Delta t < 2 \times 10^{-5} \tau$, so as to prevent the numerical instability.
Combining this aspect with the time-consuming Ewald summation, it turns out that the simulations become computationally unfeasible for investigating the flocking state transition at small dipolar coupling strengths ($\lambda < 3$).
%

%
A behavior similar to the ``micro-flocking'' state has been previously observed in systems of active Brownian disks with polar alignment. \cite{Martin-Gomez2018}
Indeed, the density $\Phi = 0.23$ considered here is close to the effective packing fraction $\phi^* = 0.256$ defined in ref. \citen{Martin-Gomez2018}, where a comparable state with `microscopic' polar clusters has been discovered at high motilities and intermediate polar coupling. 
Upon further increase in polar coupling, the system in ref. \citen{Martin-Gomez2018} displays macroscopic structures, such as moving patterns of bands or lanes.
Here, we did not find such patterns in the parameter range explored.
In this context, we also mention a state observed in the Vicsek model at low densities and low noise (yielding strong alignment). \cite{Vicsek1995}
\begin{figure}[!htbp]
  \centering
  \includegraphics[width=0.9\linewidth]{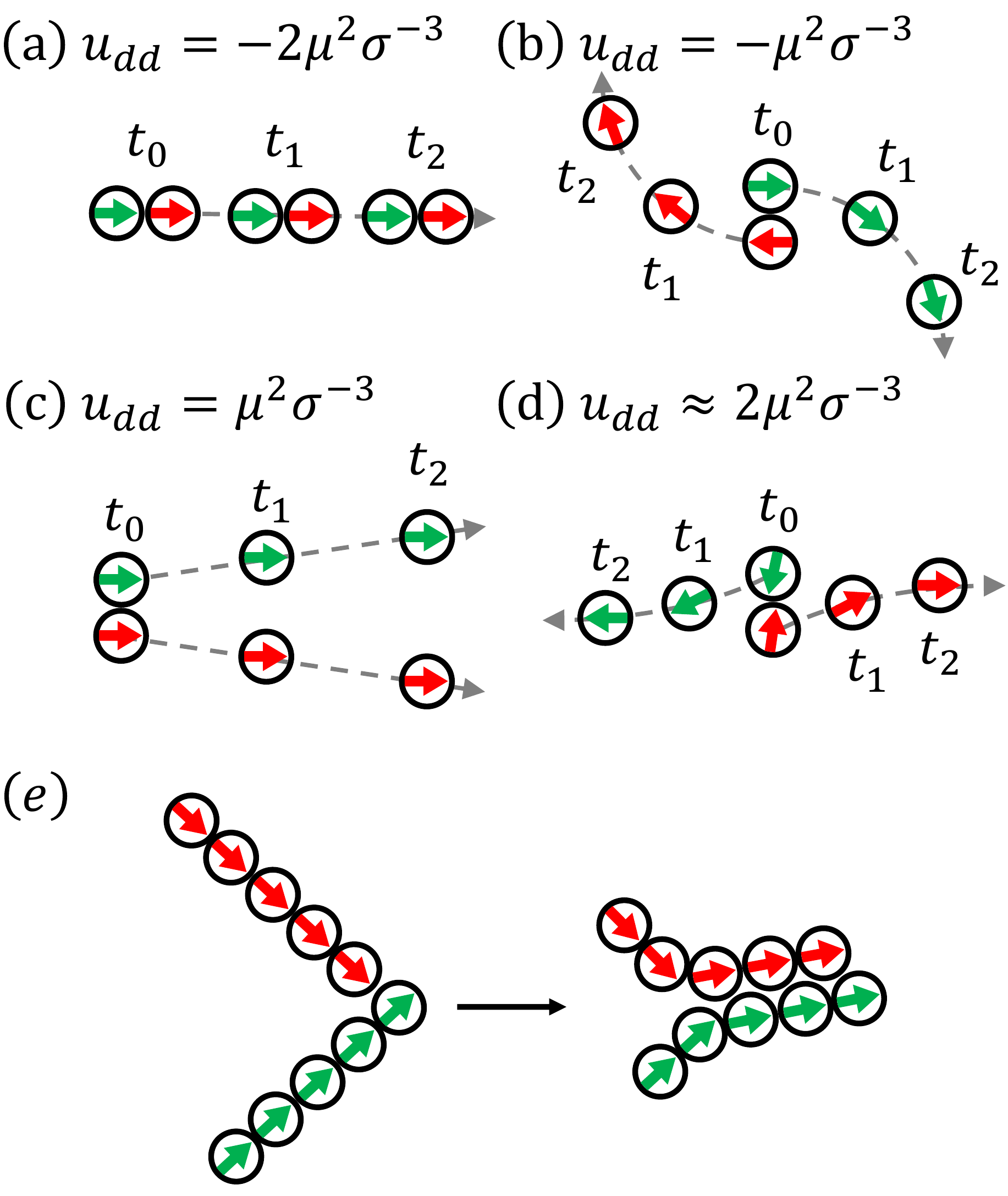}
\caption[Possible configurations at $\Phi = 0.23$]{(Color online) (a-d) Sketches of the motion of two dipolar active particles in three time steps ($t_0 < t_1 < t_2$) starting from different representative configurations. The dashed lines indicate the trajectories of the particles. (e) A scheme of a collision event of two short chains, consisting of aligned dipolar active particles.}
\label{fig:arguments}
\end{figure}
In this state, the point-like particles form small groups within which the particles move together along one (random) direction.
However, the system does not show global ordering. 
In both of these models, \textit{i.e.}, active polar disks and the Vicsek model, the particles favor alignment of the propulsion direction with their neighbors, independent of their relative positions.
This is different from the particles considered here whose interaction depends not only on the relative alignment but also on their configuration. 
%

%
In order to gain a deeper insight into the flocking behavior of the present system, we draw in Fig.~\ref{fig:arguments}(a-d) sketches of the motion of two highly motile, strongly coupled particles starting from four representative configurations. 
The initial configurations (related to time $t_0$) in Fig.~\ref{fig:arguments}(a) and (b) are those with the lowest dipolar interaction energy.
These configurations are therefore stable (up to thermal fluctuations) in the passive case. 
Upon switching on the motility ($v_0^* > 0$), the particles in the head-to-tail configuration [see Fig.~\ref{fig:arguments}(a)] will move along a straight line in the same direction (as already argued in ref. \citen{Kaiser2015}), until thermal fluctuations set in.
We thus consider this configuration as relatively stable not only in the passive, but also in the active case.
This is different in the antiparallel side-by-side configuration depicted in Fig.~\ref{fig:arguments}(b) at $t_0$.
Here, activity leads to motion in opposite direction, yielding this configuration unstable.
In Fig.~\ref{fig:arguments}(c), the initial configuration at $t_0$ is energetically unstable. 
Still, at least for short time, one expects the particles to move together along the same direction until the dipolar repulsion drives them away from one another.
Finally, in Fig.~\ref{fig:arguments}(d) the initial configuration is a slightly distorted (induced by rotational noise) head-to-head configuration.
This leads to a strong dipolar torque which, combined with strong activity, pushes the particles apart.
Combining these arguments, we conclude that the most stable configuration for two dipolar active particles is the head-to-tail alignment.
While this is similar to the passive case, a major difference is that the antiparallel side-by-side configuration [Fig.~\ref{fig:arguments}(b)] breaks immediately apart when the particles become active, yielding this configuration \textit{unstable}.
At low densities, we therefore expect to find short straight chains moving along their individual long axis, consistent with the observation in the Fig.~\ref{fig:P01Snapshots}(d).
Upon an increase in the density, short chains collide with each other more frequently than at low densities. 
As explained above, antiparallel alignment of the short chains is not stable.
Therefore, once two short chains collide, they tend to align along the same direction and, thus, form a dynamical polar cluster, as illustrated in Fig.~\ref{fig:arguments}(e).
At a sufficiently high density (\textit{e.g.}, $\Phi \gtrsim 0.23$), the frequent collisions between short chains finally lead to a ``micro-flocking'' state, in which the dipolar active particles form finite-sized clusters with a polar order [see Fig.~\ref{fig:P02Snapshots}(d)].
This is fundamentally different from the alignment mechanism in other models such as the ones with ferromagnetic-like interactions, where the configuration at $t_0$ in Fig.~\ref{fig:arguments}(c) is kinetically stable. \cite{Martin-Gomez2018, Sese-Sansa2018}
\begin{figure}[!htbp]
  \centering
  \includegraphics[width=0.9\linewidth]{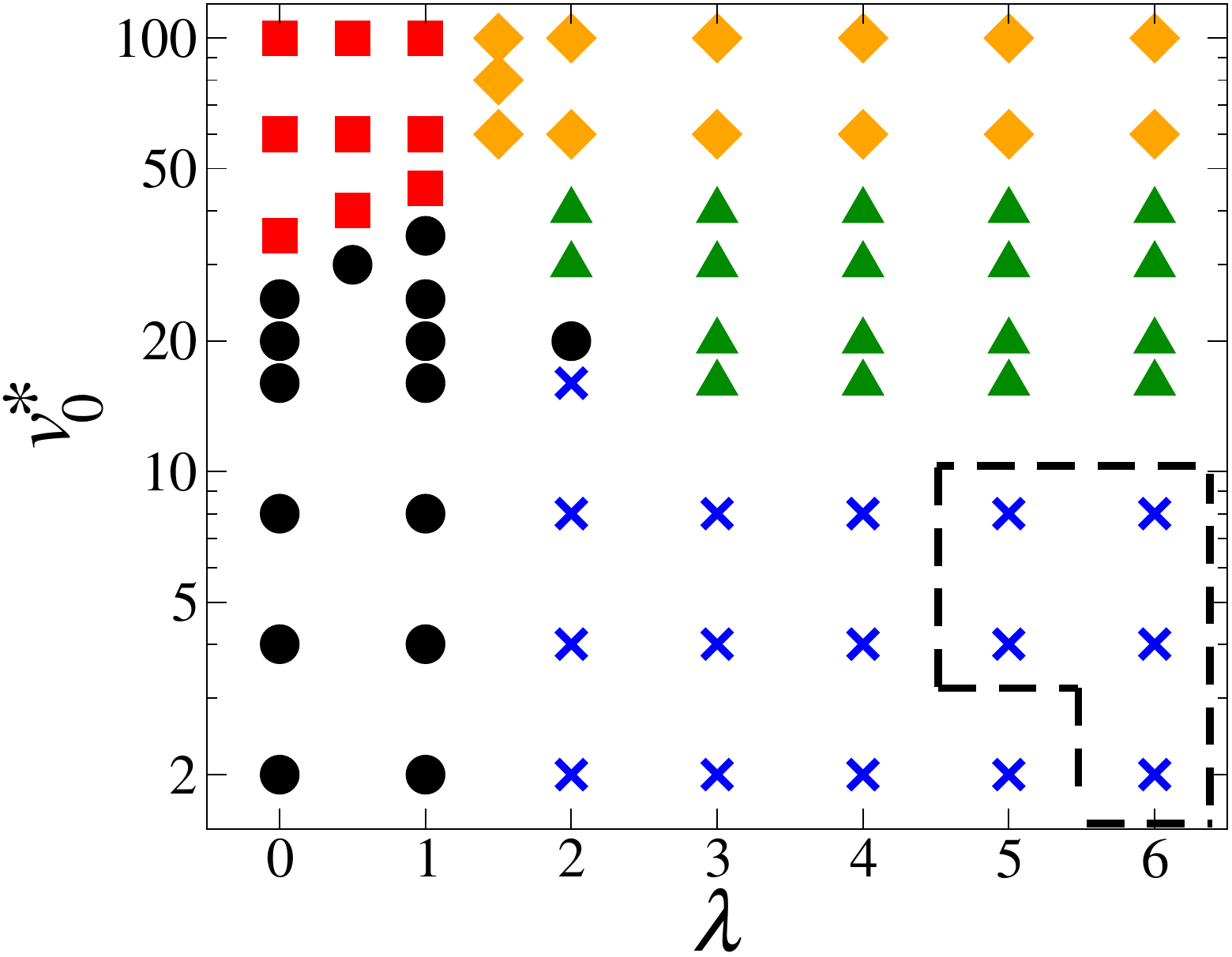}
  \caption[The state Diagram at $\Phi = 0.58$]{(Color online) State diagram of dipolar active particles in the ($v_0^*$, $\lambda$) plane at $\Phi = 0.58$. The points on the diagram indicate the parameter combinations used in the simulations. At $\Phi = 0.58$, we have observed homogeneous, isotropic fluid states (black circles), motility-induced clustering (red squares), macro-flocking (orange diamonds), micro-flocking (green triangles), and chain-like structures (blue crosses). The region surrounded by the dashed lines indicates a parameter regime where the simulations did not reach a steady state.}
  \label{fig:P05StateDiagram}
\end{figure}
\begin{figure}[!htbp]
  \centering
  \includegraphics[width=1.0\linewidth]{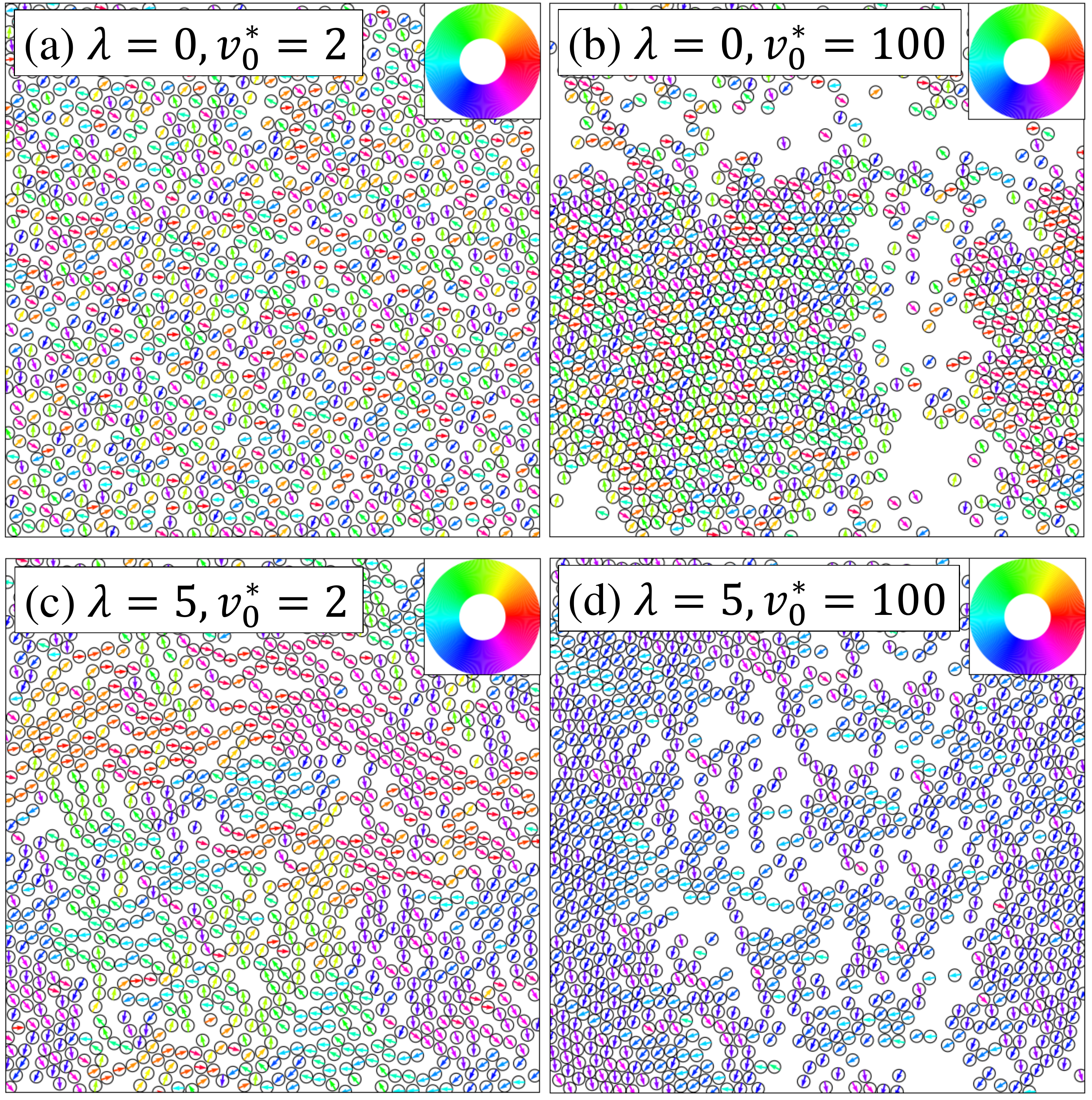}
  \caption[Snapshots at $\Phi = 0.58$]{(Color online) Representative simulation snapshots at $\Phi = 0.58$. Particles are colored according to their orientations as indicated by the color ring in the inset.}
  \label{fig:P05Snapshots}
\end{figure}

\subsection{\label{sec:clustering}The high density regime ($\Phi = 0.58$)}

%
At high densities such as $\Phi = 0.58$, there are three major phenomena interfering with one another: motility-induced phase separation, polar ordering, and chain formation.
In the following Sec.~\ref{sec:cc} $-$~\ref{sec:ch}, we discuss these issues in detail.

\subsubsection{\label{sec:cc}Motility-induced clustering}

%
Figure~\ref{fig:P05StateDiagram} shows the state diagram at $\Phi = 0.58$ in the plane spanned by motilities $v_0^*$ and dipolar coupling strengths $\lambda$. 
We start by investigating the regime of small $\lambda$.
Specifically, we are interested in the impact of dipolar interactions on the motility-induced phase separation known from conventional (non-dipolar) active particles.
Indeed, as the motility $v_0^*$ increases from zero, non-dipolar active Brownian particles ($\lambda = 0$) undergo a transition from a homogeneous, isotropic fluid state into a state with large, dense clusters coexisting with freely moving colloids in the dilute region, as shown in Fig.~\ref{fig:P05StateDiagram},~\ref{fig:P05Snapshots}(a), and (b).
This behavior, generally known as motility-induced phase separation, occurs even if attractive interactions are absent in the model. \cite{Bialke2013, Cates2015} 
The motility-induced phase separation can be explained by a self-trapping mechanism: 
When a highly motile particle enters into a dense region occupied with other particles, this particle is temporally slowed down due to frequent collisions. 
Such a slowing-down effect makes this region even denser, thus creating a positive feedback loop which leads to the formation of giant clusters. \cite{Cates2015} 
We note that the particle orientations in the two phases are essentially uncorrelated for non-dipolar particles ($\lambda = 0$), as seen in Fig.~\ref{fig:P05Snapshots}(b).
%

\begin{figure}[!htbp]
  \centering
  \includegraphics[width=0.85\linewidth]{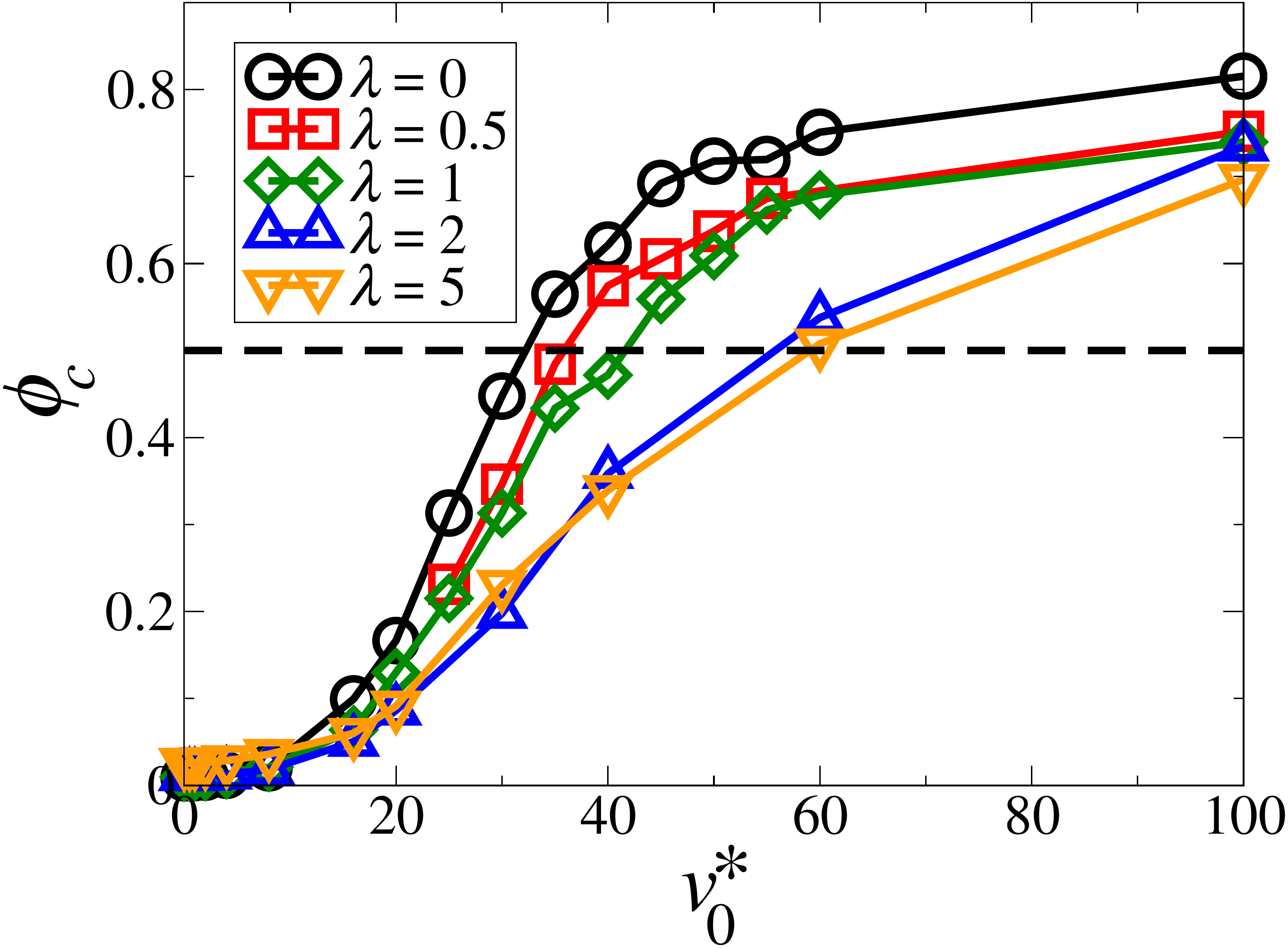}
  \caption[The clustering behavior at $\Phi = 0.58$]{(Color online) Fraction of the largest cluster $\phi_c$ as a function of the motility $v_0^*$ for the coupling strength $\lambda = 0$ (black dots), $0.5$ (red squares), $1$ (green diamonds), $2$ (blue triangles up), and $5$ (orange triangles down). The dashed line $\phi_c = 0.5$ marks the value corresponding to cluster formation.}
  \label{fig:clustering}
\end{figure}

%
To get a first impression of the impact of dipole-dipole interactions on the clustering behavior, we plot in Fig.~\ref{fig:clustering} the fraction of the largest cluster, $\phi_c$ [see eqn~\eqref{eqn:phi_c}], as a function of the motility $v_0^*$ for various coupling strengths $\lambda$. 
Consistent with the motility-induced phase separation, the curves for non-dipolar and weakly coupled active particles ($\lambda = 0 - 1$) display a sharp increase from zero at $v_0^* \approx 20$ and reach large values ($\phi_c \approx 0.8$) at $v_0^* = 100$.
The curves reach $\phi_c = 0.5$ at larger $v_0^*$ with increasing $\lambda$, which is also reflected by the boundary between black circles and red squares in Fig~\ref{fig:P05StateDiagram}.
For strong dipolar coupling ($\lambda \gtrsim 2$), the values of $\phi_c$ at high motilities remain in the same range ($\phi_c \approx 0.8$, indicating again formation of large clusters), but the increase in $\phi_c$ at small $v_0^*$ is much less pronounced.
This already indicates a strong impact of dipolar interactions.
For more detailed analysis of the clustering behavior, we discuss the cluster size distribution in Appendix~\ref{sec:clusterDistri}.

\begin{figure}[!htbp]
  \centering
  \includegraphics[width=0.85\linewidth]{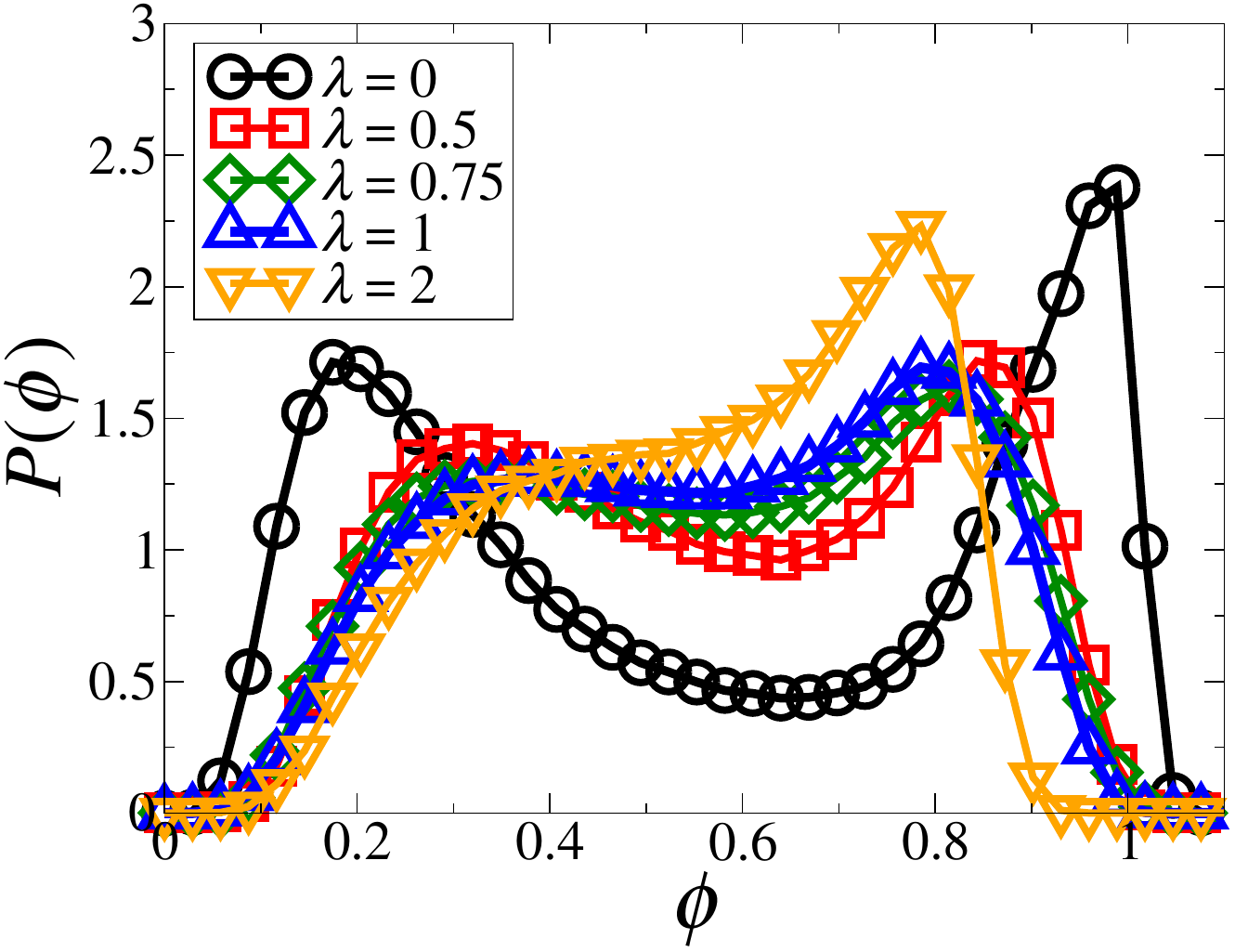}
  \caption[PDF of local area fraction at $\Phi = 0.58$]{(Color online) Probability distribution function of local area fractions, $P\left(\phi\right)$, for the coupling strength $\lambda = 0$ (black dots), $0.5$ (red squares), $0.75$ (green diamonds), $1$ (blue triangles up), and $2$ (orange triangles down) with the motility $v_0^* = 100$.}
  \label{fig:PDF_localphi}
\end{figure}

%
To gain further information, we calculate position-resolved local area fractions, $\phi$, based on a Voronoi tessellation. \cite{Blaschke2016, Liao2018}
(We note that, contrary to ref. \citen{Liao2018}, we did not perform a short time average of $\phi$, since the polar clusters characterizing the dense state migrate over time and, therefore, are not stationary within the short-time interval.
The flocking behavior will be discussed later in detail in Sec.~\ref{sec:fl}.)
Figure~\ref{fig:PDF_localphi} shows the probability distribution of the local area fractions, $P\left(\phi\right)$, for different coupling strengths at $v_0^* = 100$.
For non-dipolar active particles ($\lambda = 0$), $P\left(\phi\right)$ reveals a clear double-peak structure that reflects the separation between the dilute and the dense phase.
\begin{figure}[!htbp]
  \centering
  \includegraphics[width=0.85\linewidth]{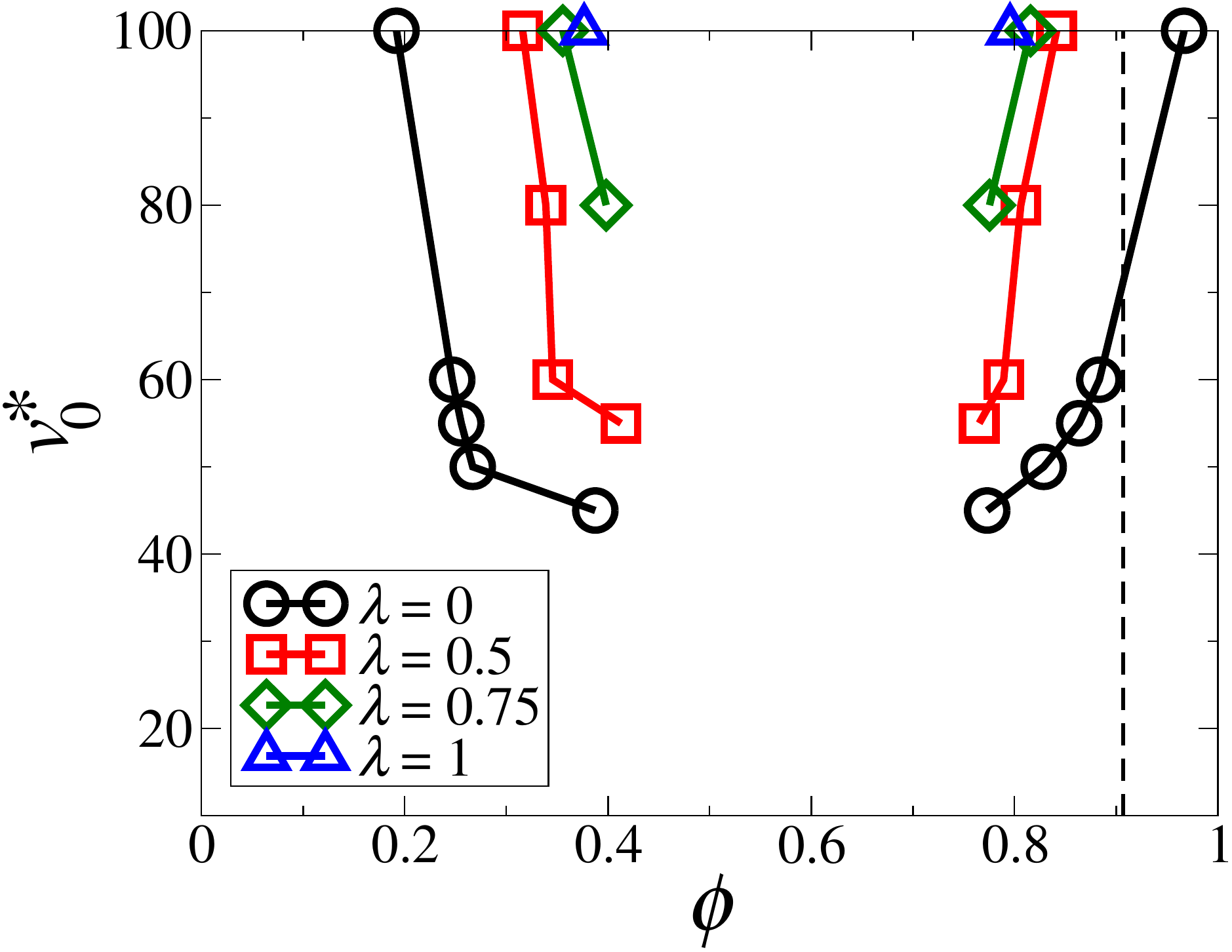}
  \caption[Area fractions corresponding to coexisting states at $\Phi = 0.58$]{(Color online) Coexisting densities in the $\left(v_0^*, \phi \right)$ plane for the coupling strength $\lambda = 0$ (black dots), $0.5$ (red squares), $0.75$ (green diamonds), and $1$ (blue triangle up). The black dashed line displays the effective close-packing fraction, $\phi_{cp} = \pi /\left(2 \sqrt{3}\right) \approx 0.91$.}
  \label{fig:phaseSep}
\end{figure}
The coexisting densities correspond to the location of the two local maxima of $P\left(\phi\right)$.
Upon an increase in the coupling strength up to $\lambda = 2$, the double-peak structure gradually disappears. 
Instead, we observe the emergence of a single peak located at a density slightly larger than the mean density ($\Phi = 0.58$), as well as a broad shoulder on the left.
This suggests the disappearance of the phase separation observed at zero and small $\lambda$.
%

%
The same conclusion can be drawn from Fig.~\ref{fig:phaseSep}, where we plot the coexisting densities in the ($v_0^*$, $\phi$) plane for various $\lambda$.
The location of the high-density branch at $\lambda = 0$ can be explained as follows:
At high motilities $v_0^*$, the non-dipolar system ($\lambda = 0$) exhibits ``giant'' clusters, composed of randomly oriented particles [see Fig.~\ref{fig:P05Snapshots}(b)].
These particles are separated by a distance close to the effective hard sphere diameter $\sigma_{eff}$ defined in Sec.~\ref{sec:BD}.
As a result, the local area fraction in the densely packed region should be close to the close-packing fraction, $\phi_{cp} = \pi /\left(2 \sqrt{3}\right) \approx 0.91$.
Upon a slight increase in dipolar coupling, the coexistence branches for the dilute region, $\phi_{gas}$, are shifted toward higher densities, while the branches for the dense region, $\phi_{den}$, move toward lower area fractions.
As a result, the area surrounded by the curves of $\phi_{den}$ and $\phi_{gas}$ in Fig.~\ref{fig:phaseSep} significantly shrinks with increasing $\lambda$, indicating that motility-induced phase separation is generally suppressed by the dipolar interactions.
Within the present simulations, the phase separation and the corresponding coexistence curves disappear once $\lambda > 1$.
We note that this observation is in contrast with the findings in a recent study of active Brownian particles with additional interactions of velocity alignment at high densities. \cite{Sese-Sansa2018}
This shows that different types of orientational interactions may have entirely different impacts on motility-induced phase separation. 
%

\begin{figure}[!htbp]
  \centering
  \includegraphics[width=0.85\linewidth]{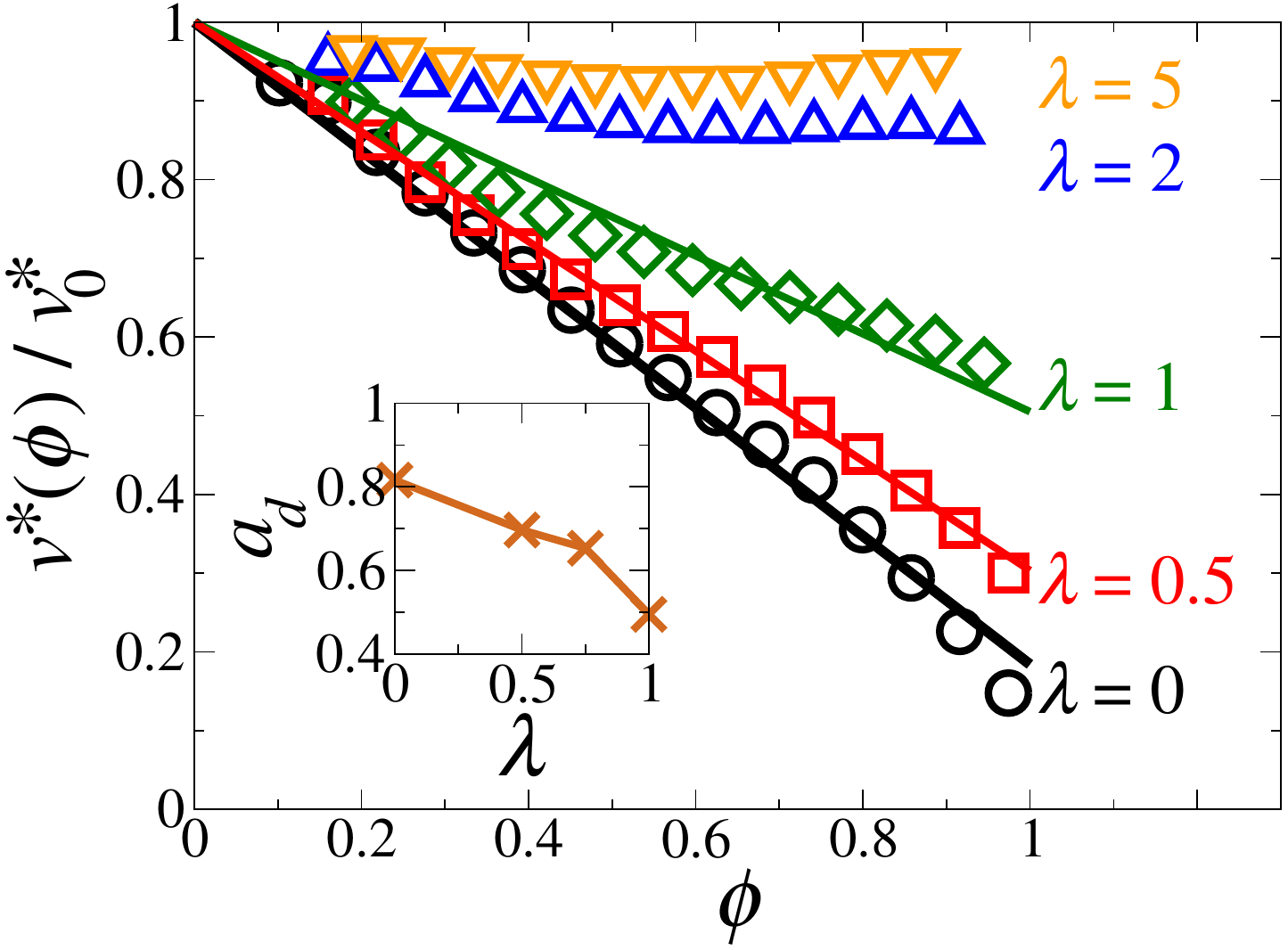}
  \caption[speed of phi at $\Phi = 0.58$]{(Color online) Normalized particle speed $v^*\left(\phi\right)/ v_0^*$ as a function of the local area fraction $\phi$ for the coupling strength $\lambda = 0$ (black dots), $0.5$ (red squares), $1$ (green diamonds), $2$ (blue triangles up), and $5$ (orange triangles down) with the motility $v_0^* = 100$. The solid lines represent fits to eqn~\eqref{eqn:linear_decay}. Inset: Fitted decay amplitude $a_d$ as a function of $\lambda$. The line is drawn as a guide to the eye.}
  \label{fig:vec_of_phi}
\end{figure}

%
Yet another perspective on the disappearance of the phase separation due to dipolar interactions emerges when we consider the self-trapping mechanism (which plays a key role for clustering at $\lambda = 0$).
To this end, we compute the normalized speed of the $i$th particle, $v_i^* / v_0^*$, \textit{versus} the local density $\phi_i = \phi\left(\boldsymbol{r}_i\right)$, where the speed is given by $ v_i^* = \vert \left( \Delta\boldsymbol{r}_i / \sigma \right) / \left( \Delta t_s / \tau \right) \vert$ with $\Delta t_s = 10^{-2} \tau$.
After averaging over $N$ particles and over at least $1000$ snapshots, we plot in Fig.~\ref{fig:vec_of_phi} the normalized particle speed \textit{versus} the local density.
For non-coupled ($\lambda = 0$) and weakly coupled ($0 < \lambda \lesssim 1$) systems, we observe a linear decay,
\begin{equation} \label{eqn:linear_decay}
  v^*\left(\phi\right)/v_0^* = 1 - a_d \phi\text{,}
\end{equation}
consistent with the prediction from a phenomenological approach: \cite{Stenhammar2013, Stenhammar2014} 
Particles move slower when traveling through a crowded area.
The fitting parameter, $a_d$, represents the decay amplitude, and, therefore, describes the significance of the self-trapping mechanism. 
Upon increasing $\lambda$ in the range $\lambda \lesssim 1$, the decay becomes less pronounced.
Consequently, the fitting parameter, $a_d$, monotonically decreases, as seen in the inset of Fig.~\ref{fig:vec_of_phi}.
In contrast, for $\lambda \gtrsim 2$, the normalized speed as a function of $\phi$ is essentially constant, that is, $a_d$ tends to zero.
In other words, there is no self-trapping anymore.
This may be attributed to the fact that the strongly-coupled dipolar particles tend to form polar clusters with local head-to-tail alignment, causing the particles to move along the same direction [see Fig.~\ref{fig:P05Snapshots}(d)].
As a result, the motility-induced phase separation is replaced by a macro-flocking state, as shown in Fig.~\ref{fig:P05StateDiagram}.
We will come back to the flocking behavior in Sec.~\ref{sec:fl}.

%
To obtain a more complete (yet qualitative) picture on the role of the function $a_d\left(\lambda\right)$ for the appearance of the motility-induced phase separation, we consider the effective free energy proposed in ref.~\citen{Tailleur2008}, 
\begin{equation}
  f\left(\phi\right) = f_0\left(\phi\right) + f_{rep}\left(\phi\right)\textit{,}
\end{equation}
where the bulk contribution is given by
\begin{equation}
  f_0\left(\phi\right) = \phi \left(\text{ln}\phi - 1 \right) + \int_{0}^{\phi} \text{ln} \left[v\left(\rho\right)\right] d \rho
\end{equation}
with density-dependent swim speed $v\left( \rho \right) = 1 - a_d \rho$. \cite{Stenhammar2013, Stenhammar2014} 
Further, the contribution from the excluded volume interactions between particles is written as 
\begin{equation}
  f_{rep}\left(\phi\right) = k_{rep} \Theta\left(\phi - \phi_t\right)\left(\phi - \phi_t\right)^4
\end{equation}
with $\Theta\left(x\right)$ being the Heaviside step function, $k_{rep}$ the repulsive strength, and $\phi_t$ the threshold area fraction.
%

\begin{figure}[!htbp]
  \centering
  \includegraphics[width=0.9\linewidth]{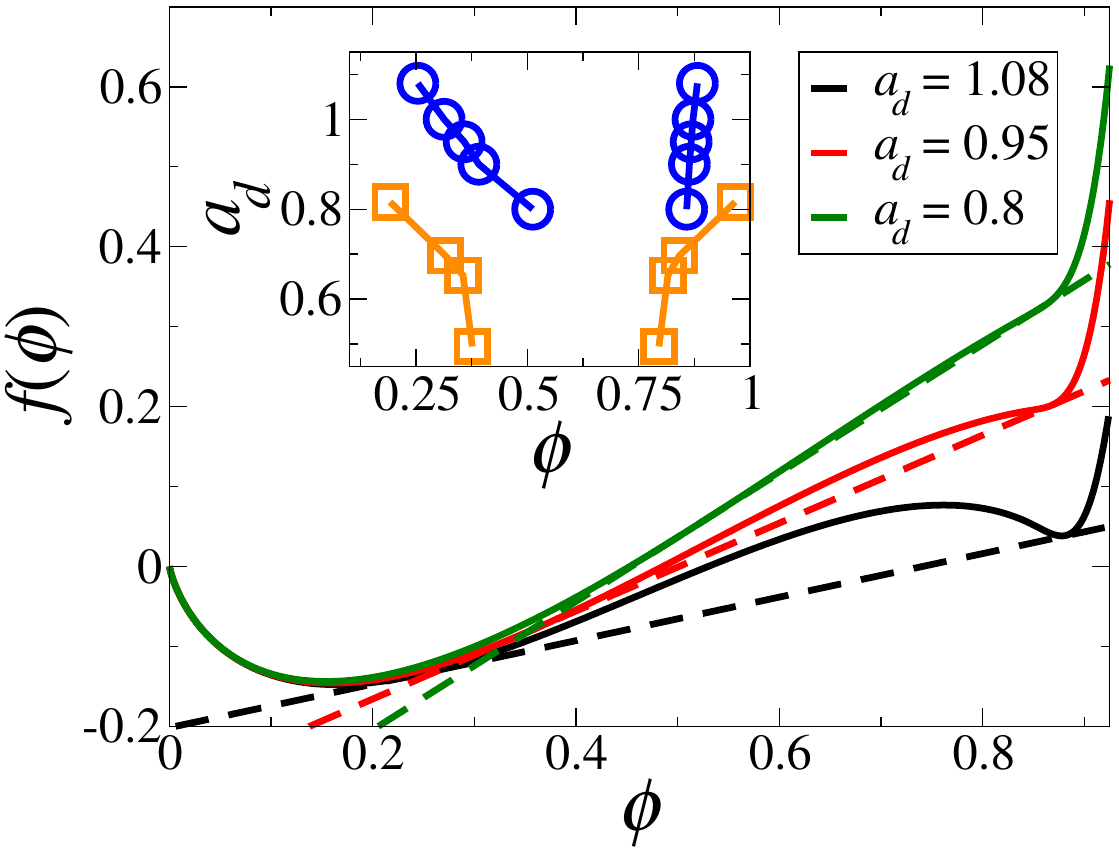}
  \caption[argument from the perspective of the effective free energy]{(Color online) Effective free energy $f\left(\phi\right)$ with $k_{rep} = 5000$ and $\phi_t = 0.84$ for various $a_d = 1.08$ (solid black line), $0.95$ (solid red line), and $0.8$ (solid green line). The dashed lines display the common tangent construction. Terms linear in $\phi$ are irrelevant for the common tangent construction and have been subtracted for clarity. Inset: Dependence of coexisting densities on $a_d$ obtained from the effective free energy (blue circles) and simulations at $v_0^* = 100$ and $0 \leq \lambda \leq 1$ (orange squares).}
  \label{fig:eff_free_energy}
\end{figure}

%
For conventional active Brownian particles ($\lambda = 0$), the decay amplitude, as a crucial parameter describing the degree of self-trapping, is typically chosen to be $a_d = 1$ for 2D and $a_d = 1.3$ for 3D. \cite{Stenhammar2014}
With this in mind, in Fig.~\ref{fig:eff_free_energy} we plot the effective free energy $f\left(\phi\right)$ for the decay amplitude $a_d = 1.08$, $0.95$, and $0.8$.
At low densities, $f\left(\phi\right)$ is nearly independent of $a_d$, while at high densities, $f\left(\phi\right)$ increases with decreasing $a_d$. 
From $f\left(\phi\right)$, we determine the coexisting densities through the common tangent construction.
To compare the influence of the decay amplitude, $a_d$, on the phase separation from the theoretical perspective and the simulation results, we plot in the inset of Fig.~\ref{fig:eff_free_energy} the coexisting densities in the $\left( a_d , \phi \right)$ plane for the effective free energy (blue circles) and simulations at $v_0^* = 100$ and $0 \leq \lambda \leq 1$ (orange squares).
More specifically, for the above simulations we plot the coexisting densities determined from Fig.~\ref{fig:PDF_localphi} versus the associated $a_d$ plotted in the inset of Fig.~\ref{fig:vec_of_phi}.
The coexisting densities obtained from both methods describe the same trend, that is, a decrease in the decay amplitude $a_d$ causes a reduction in $\phi_{den}$ and a growth in $\phi_{gas}$.
Further, the phase separation disappears once $v\left(\phi\right)$ decays sufficiently slow ($a_d \lesssim 0.6$ for the effective free energy, and $a_d \lesssim 0.5$ for simulations).
This clearly indicates that the motility-induced phase separation is suppressed when the function $v\left(\phi\right)$ decreases too slowly, \textit{i.e.}, when the dipolar coupling becomes too strong.

\subsubsection{\label{sec:fl}Flocking}

\begin{figure}[!htbp]
  \centering
  \includegraphics[width=0.8\linewidth]{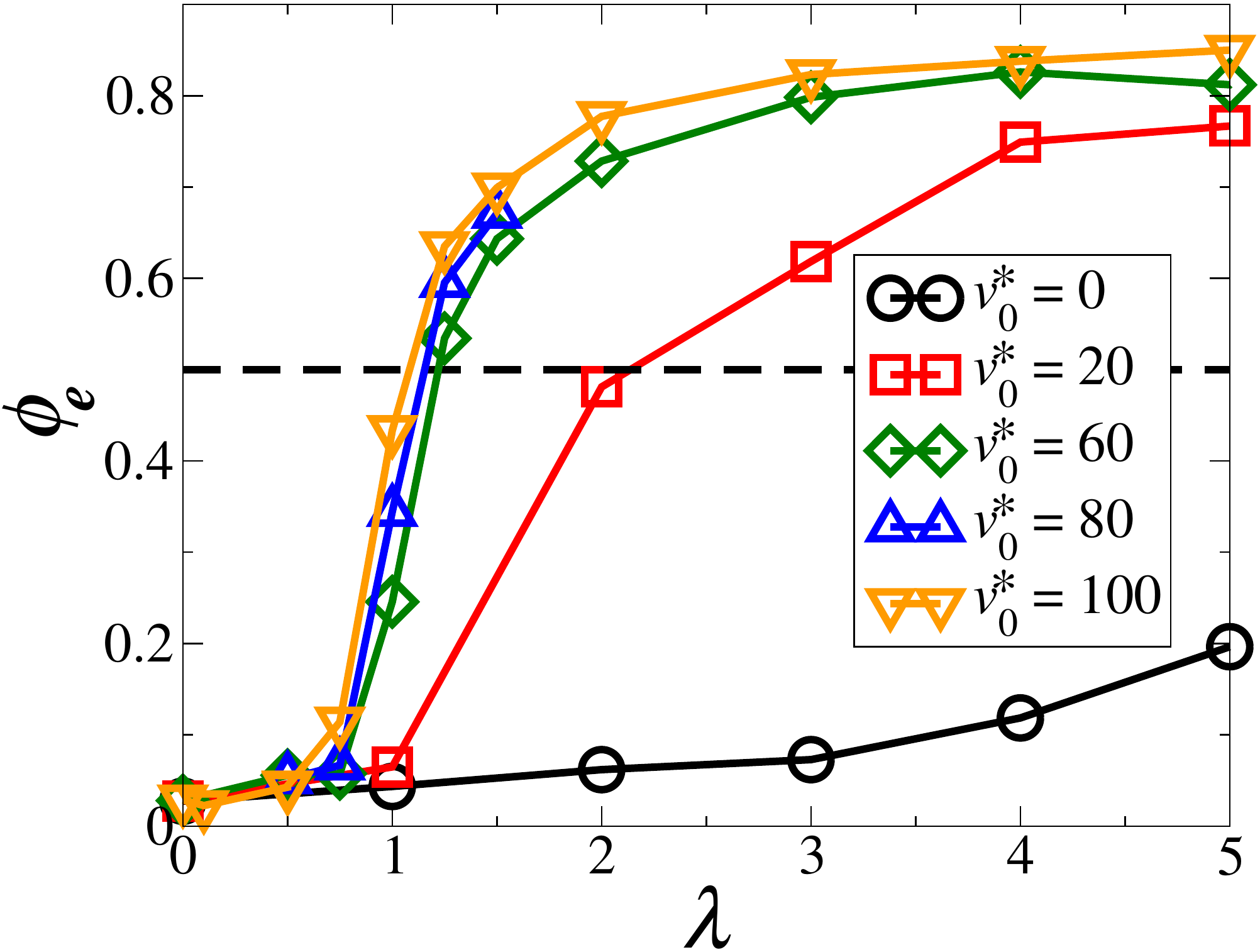}
  \caption[Magnitude of the average orientation, $\phi_{\boldsymbol{e}}$, at $\Phi = 0.58$]{(Color online) Magnitude of the average orientation $\phi_{\boldsymbol{e}}$ as a function of the coupling strength $\lambda$ at $\Phi = 0.58$ for $v_0^* = 0$ (black dots), $20$ (red squares), $60$ (green diamonds), $80$ (blue triangles up), and $100$ (orange triangles down). The dashed line $\phi_{\boldsymbol{e}} = 0.5$ indicates the value above which the particles form a flocking state (according to Table~\ref{table:states}). The solid lines are drawn as a guide to the eye.}
  \label{fig:P05phi_e}
\end{figure}

%
We now come back to the emergence of flocking and polar ordering (see green triangles and orange diamonds in Fig.~\ref{fig:P05StateDiagram}).
As argued in Sec.~\ref{sec:ordering}, once the density is sufficiently high ($\Phi \gtrsim 0.23$), the interplay of dipolar interactions and activity allows for the formation of polar clusters.
To quantify this behavior at high densities, Fig.~\ref{fig:P05phi_e} shows the magnitude of the average orientation, $\phi_{\boldsymbol{e}}$, as a function of $\lambda$ for various values of $v_0^*$.
In the passive case ($v_0^* = 0$), the order parameter remains small for all $\lambda$ considered.
Thus, there is no clear hint regarding whether global order appears, which is consistent with earlier simulations of monolayers of dipolar particles. \cite{Weis2002, Weis2003, Luo2009}
From the simulations, it is known that dense two-dimensional passive systems of dipolar particles tend to develop domain-like structures, which are highly frustrated and characterized by the polar order only locally.
Coming back to the active case, we see from Fig.~\ref{fig:P05phi_e} that already at an intermediate value of the motility ($v_0^* = 20$), the order parameter reaches significant values ($\phi_e \gtrsim 0.5$) when $\lambda$ exceeds a value of about two. 
Combining this finding with the cluster analysis shown in Fig.~\ref{fig:clustering} and using the criteria in Table~\ref{table:states}, we classify this behavior as a micro-flocking state ($3 \lesssim \lambda \lesssim 6$, $10 \lesssim v_0^* \lesssim 40$), see green triangles in Fig.~\ref{fig:P05StateDiagram}.
Finally, in the regime of high motilities ($v_0^* = 60 - 100$), the data curve $\phi_{\boldsymbol{e}}\left(\lambda\right)$ is reminiscent of a (polar) phase transition:
$\phi_{\boldsymbol{e}}$ rises suddenly from zero to values greater than $0.5$ at a ``critical'' coupling strength $\lambda_c \approx 0.75$.
This critical value slightly decreases upon an increase in $v_0^*$.
In addition to the large values of $\phi_{\boldsymbol{e}}$, the fraction of the largest cluster, $\phi_c$, is greater than $0.5$ for $v_0^* \gtrsim 60$, as shown in Fig.~\ref{fig:clustering}.
Thus the systems at $v_0^* \gtrsim 60$ and $\lambda \gtrsim 1.25$ are in a macro-flocking state. 
We note that the existence of the polar order is also reflected by the positive time correlations of individual dipole moments at a time interval much longer than the Brownian diffusion time $\tau$ (not shown here).

\subsubsection{\label{sec:ch}Chain-like structures}

%
Having discussed the emergence and interplay of phase separation and global polar order, we finally consider the fate of the chain-like structures characterizing passive, dense dipolar system upon increasing $v_0^*$ from zero.
The question is how the system transforms from a state with chain-like structures into a macro-flocking state [see Fig.~\ref{fig:P05StateDiagram},~\ref{fig:P05Snapshots}(c), and~\ref{fig:P05Snapshots}(d)].
As discussed in Sec.~\ref{sec:chain}, the order parameter $\phi_p$ (degree of polymerization) is no longer adequate to quantify such a transition at high densities.
\begin{figure}[!htbp]
  \centering
  \includegraphics[width=0.8\linewidth]{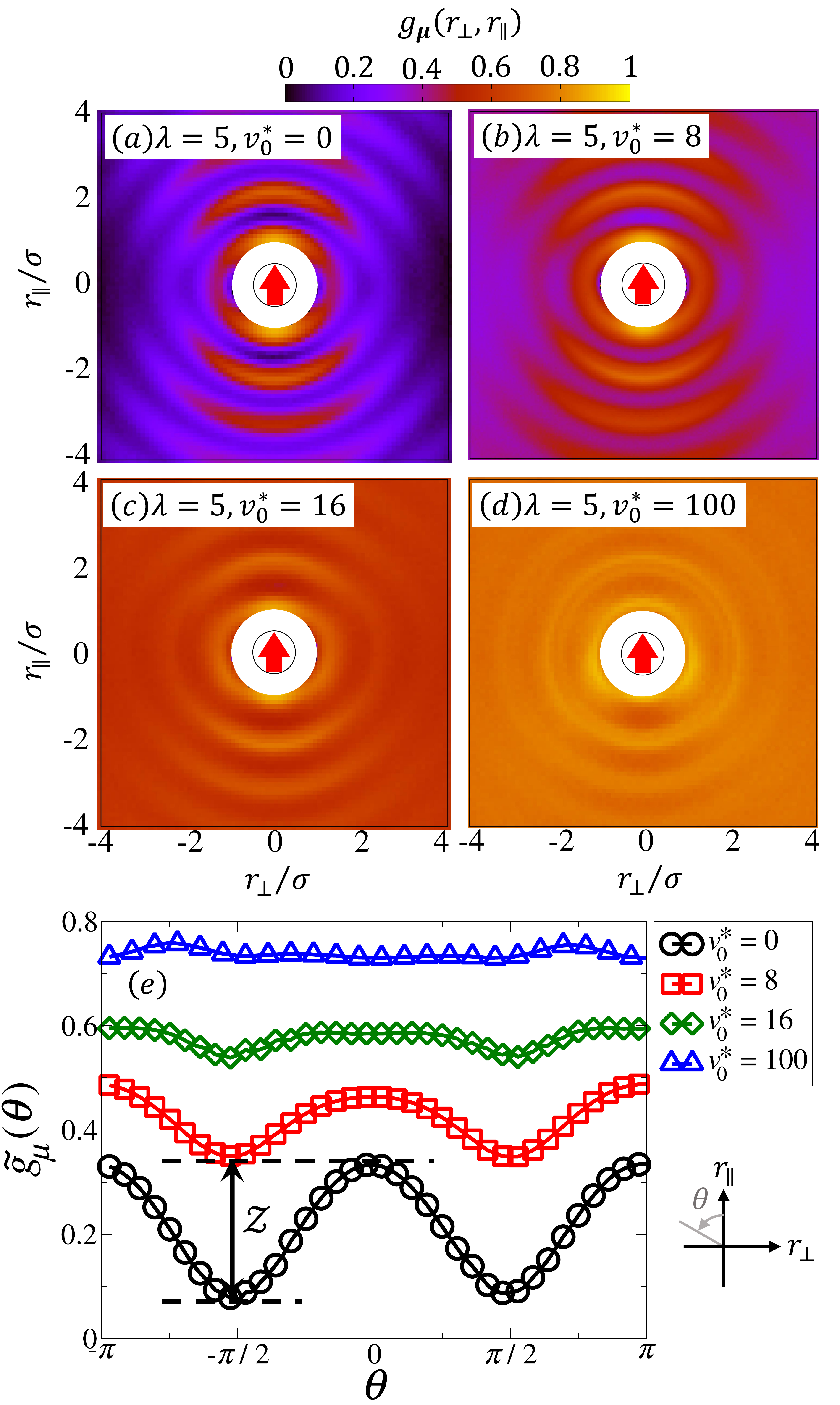}
  \caption[Correlation functions of dipole moments at $\Phi = 0.58$]{(Color online) Spatial correlation function of dipole moments, $g_{_{\boldsymbol{\mu}}}\left( r_{\bot}, \: r_{\|} \right)$ defined in eqn~\eqref{eqn:g_mu_xy}, at $\Phi = 0.58$ and $\lambda = 5$ for the motility $v_0^* = 0$ (a), $8$ (b), $16$ (c), and $100$ (d). The black circle represents the reference particle with the red arrow indicating the particle's orientation as well as the dipole moment. The area $r = \sqrt{r_{\bot}^2 + r_{\|}^2} < \sigma$ is drawn in white color to reflect that the center-to-center distance between particles can not be smaller than its diameter due to steric repulsion. (e) The angular correlation function of dipole moments, $\tilde{g}_{_{\boldsymbol{\mu}}}\left(\theta \right)$ defined in eqn~\eqref{eqn:g_mu_theta}, as a function of $\theta = \text{atan2}\left(-r_{\bot}, \: r_{\|}\right)$ with the coupling strength $\lambda = 5$. The graph at the bottom-right corner of (e) illustrates the definition of $\theta$.}
  \label{fig:g_mu}
\end{figure}
\begin{figure}[!htbp]
  \centering
  \includegraphics[width=0.8\linewidth]{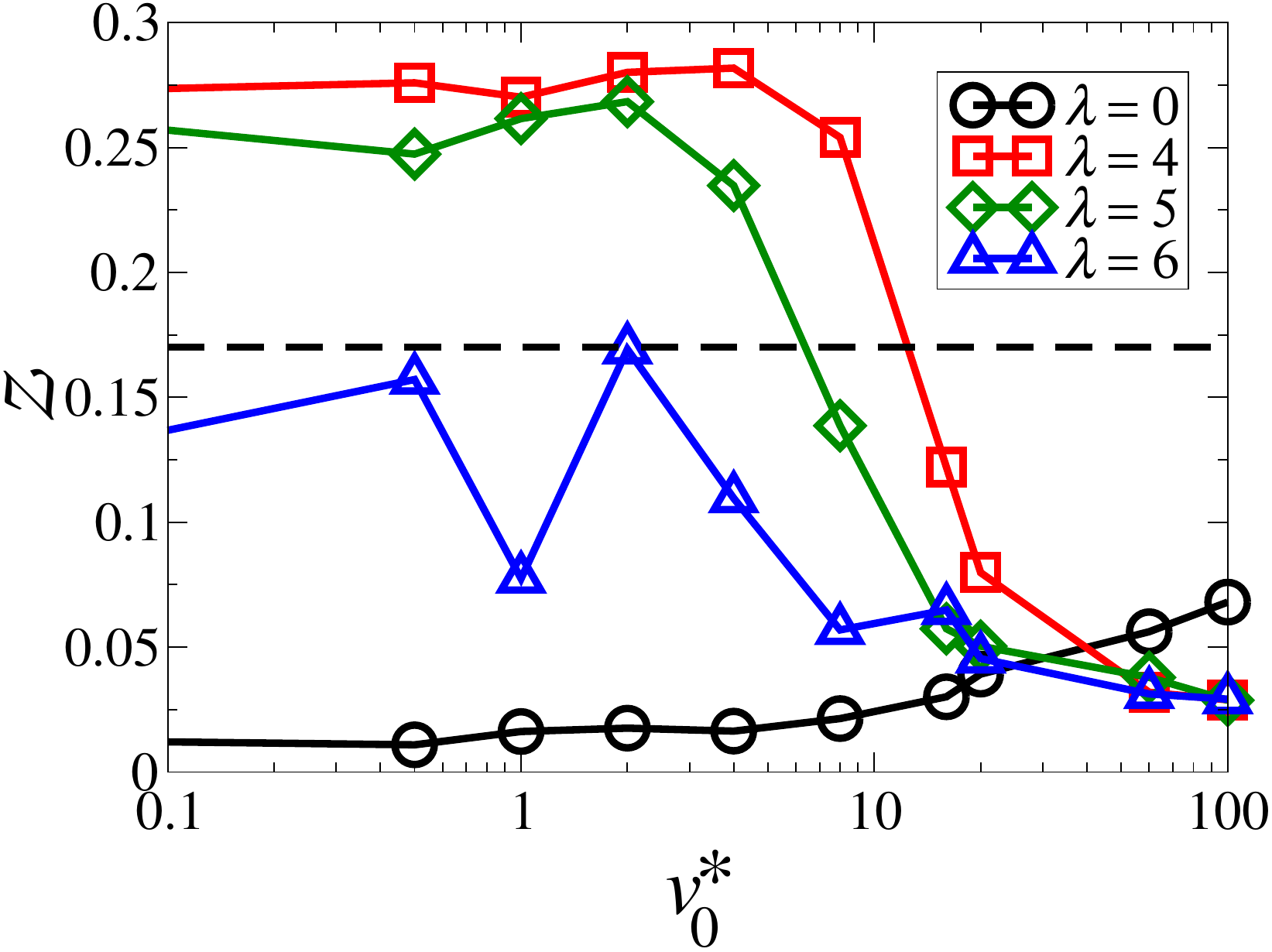}
  \caption[The order parameter $\mathzapf{Z}$ at $\Phi = 0.58$]{(Color online) Order parameter $\mathzapf{Z}$ as a function of the motility $v_0^*$ at $\Phi = 0.58$ for the coupling strength $\lambda = 0$ (black dots), $4$ (red squares), $5$ (green diamonds), and $6$ (blue triangles). The dashed line indicate the threshold value for the state transition $\mathzapf{Z}_{thres} = 0.17$. The solid lines are drawn as a guide to the eye.}
  \label{fig:zeta}
\end{figure}
Therefore, an alternative method is proposed: 
Figure~\ref{fig:g_mu}(a-d) shows the spatial correlation function of dipole moments, $g_{_{\boldsymbol{\mu}}}\big( r_{\bot}, r_{\|} \big)$ [defined in eqn~\eqref{eqn:g_mu_xy}], for $v_0^* = 0 - 100$ and $\lambda = 5$.
For the passive and nearly passive case [see Fig.~\ref{fig:g_mu}(a-b)], the orange (curved) strips clearly indicate positive correlations in front of and behind the reference particle, while the purple regions suggest that the dipole moments in the equatorial zone are rather uncorrelated. 
This observation reflects chain formation along the direction of the reference dipole moment.
As the motility $v_0^*$ increases [see Fig.~\ref{fig:g_mu}(c-d)], the purple regions disappear.
Instead, the whole figure turns orange, indicating positive correlations in all directions.
This shows a rather uniform alignment of dipole moments regardless of the relative positions between particles.
In other words, the chains disappear.
%

%
To better evaluate the dependence of orientational correlations on the direction of the connecting vector $\boldsymbol{r}_{ij}$, Fig.~\ref{fig:g_mu}(e) shows the angular correlation function, $\tilde{g}_{_{\boldsymbol{\mu}}}\left(\theta \right)$ [defined in eqn~\eqref{eqn:g_mu_theta}], for the strong dipolar coupling strength $\lambda = 5$.
As mentioned in Sec.~\ref{sec:chain}, the difference between the maximum $\tilde{g}_{_{\boldsymbol{\mu}}}\left(\theta_{max}\right)$ and the minimum $\tilde{g}_{_{\boldsymbol{\mu}}}\left(\theta_{min}\right)$ of the curve, $\mathzapf{Z}$ [defined in eqn~\eqref{eqn:Z}], can be used as a measure of chain formation.
Upon increasing $v_0^*$, $\mathzapf{Z}$ decreases and, at the same time, the curves in Fig.~\ref{fig:g_mu}(e) are shifted toward lager positive correlations.
An overview of the behavior of $\mathzapf{Z}$ as a function of $v_0^*$ (for various $\lambda$) is given in Fig.~\ref{fig:zeta}.
The data curves bear close resemblance to those for $\phi_p$ in Fig.~\ref{fig:phi_p}.
For a non-dipolar system ($\lambda = 0$), where chains are absent, the order parameter $\mathzapf{Z}$ does not show any significant behavior within the range of motilities explored.
In contrast, at large coupling strengths ($\lambda = 4$ and $\lambda = 5$), the decrease in $\mathzapf{Z}$ with increasing $v_0^*$ reflects the disappearance of a state with chain-like structures.
Specifically, the curves for $\lambda \geq 4$ show a sharp decrease, accompanied by a point of inflection $\mathzapf{Z} \approx 0.17$. 
We choose this as a threshold value.
We note that the simulations of nearly passive, strongly-coupled systems (\textit{e.g.}, $\lambda = 5$ and $6$) are severely plagued by large fluctuations of the order parameter $\mathzapf{Z}$ over time.
In these cases, we were unable to determine whether or not the simulations had reached a steady state even after a very long simulation time ($t > 300 \tau$).
The corresponding region in the state diagram in Fig.~\ref{fig:P05StateDiagram} is marked by dashed lines.

\section{\label{sec:conclusions}Conclusions}

%
Using Brownian Dynamics simulations, we have studied how dipolar interactions and self-propulsion combine to influence the dynamical self-assembly of a monolayer of dipolar active Brownian particles.
To this end, we have presented state diagrams in the plane spanned by the dipolar coupling and the motility for three representative densities.
%

%
When the motility is small, the state diagrams are similar for all densities considered.
Specifically, homogeneous and isotropic fluids are observed for nearly passive particles with weak dipolar coupling, whereas strong dipolar coupling leads to chain-like structures.
At high motilities, the state diagrams strongly depend on the mean area fraction $\Phi$.
At low densities ($\Phi = 0.12$) and strong dipolar coupling, an increase in the motility $v_0^*$ from zero causes chain-like structures to break into fragments of short chains and individual beads.
At intermediate densities, passive, strongly-coupled dipolar particles self-assemble into a state with chain-like structures.
With increasing $v_0^*$ from zero, chains start to break and the system displays a micro-flocking state, where particles form finite-sized clusters with polar order.
Finally, at a high densities and strong dipolar interactions, we observe a motility-induced transition from a state with chain-like structures into a micro-flocking state and finally into a macro-flocking state, where particles show global orientational ordering and form ``giant'' clusters.

%
To provide a simple argument for the emergence of polar order (which is absent in the passive 2D case), we have considered the time evolution of two dipolar particles starting from four representative initial configurations.
As a result of the interplay between dipolar coupling and self-propulsion, the head-to-tail configuration remains the most stable one (same as the passive case), while the antiparallel side-by-side configuration is destabilized.
With increasing particle number, the head-to-tail alignment mechanism can finally lead to a flocking state if the density is sufficiently high ($\Phi \gtrsim 0.23$).
%

%
It has been reported earlier that finite size effects may hide crucial features of the flocking behavior in active systems, such as the first order nature of the flocking transition followed by the formation of traveling bands. \cite{Chate2008}
We note that the sizes of the present simulations are limited to the order of $10^3$ particles, due to the long-range character of the dipolar interactions and, subsequently, the expensive computational cost.
Besides the qualitative behavior, it would be very interesting to study the scaling behavior and explore whether the dipolar active particles belong to any of the existing universality classes, such as the one of the Vicsek model. \cite{Toner1995, Baglietto2012, Mahault2019}
To this end, however, it would be necessary to perform extensive simulations with system sizes much larger than $N \approx 10^3$, which is, again, limited by the computational resources.
Nevertheless, it is worth mentioning a fundamental difference regarding the flocking transition: 
In the Vicsek model the onset of the flocking state is characterized by simultaneous spontaneous appearance of density and orientational inhomogeneities. \cite{Vicsek1995, Chate2008, Solon2015}
In contrast, in our model the emergence of global orientational order is decoupled from the formation of large-scale structures, such as ``giant'' clusters.
%

%
At $\Phi = 0.58$ and large values of $v_0^*$, the system displays motility-induced phase separation, where ``giant'' clusters composed of densely-packed particles with random orientations coexist with freely moving particles in the dilute region.
The phase separation persists as long as the dipolar coupling $\lambda$ is negligible against thermal fluctuations ($0 < \lambda \lesssim 1$).
Once the dipolar interactions dominate ($\lambda > 1$), the orientations of the active particles are no longer uncorrelated and the particles tend to align with their neighbors, leading to a break down of the self-trapping mechanism and a subsequent suppression of the phase separation.
%

%
We note that our model does not account for hydrodynamic interactions between the particles. 
In the absence of dipolar coupling, simulation studies have reported that hydrodynamic interactions tend to suppress motility-induced phase separation due to, either, the near-field interactions, \cite{Yoshinaga2017, Yoshinaga2018} or the rapid decorrelation of the particle orientations. \cite{Matas-Navarro2014, Matas-Navarro2015, Theers2018}
Through investigating the pair distribution function, it has been shown that hydrodynamic interactions generally damp out the translational structure of active particles at high densities. \cite{Schwarzendahl2019}
Moreover, hydrodynamically interacting particles with a certain range of force dipole strengths can spontaneously form a state with global polar order, \cite{Alarcon2013, Delmotte2015, Yoshinaga2017, Yoshinaga2018} which can be attributed to either the actively induced rotation-translation coupling, \cite{Hoell2018} or the near-field lubrication forces. \cite{Yoshinaga2017, Yoshinaga2018}
Therefore, one may expect that dipolar coupling and hydrodynamic interactions combine to further suppress the phase separation and promote the polar ordering.
The details of the resulting collective behavior are a topic of future studies.
%

%
In the real world, active particles are often asymmetric. \cite{Lauga2006, Loose2014}
As a result, the effective propulsion force does not coincide with the particle's center of mass, thus generating a propulsion torque which induces chiral active motion. \cite{VanTeeffelen2008, Kummel2013}
Such a mechanism is also expected for the magnetic or dielectric Janus particles, whose dipole moments (either permanent or induced) are mostly shifted from the center of the individual particles. \cite{Baraban2008, Steinbach2016, Yener2016, Klapp2016}
Therefore, it would be very interesting to investigate models accounting for the chiral motion and the shifted dipole moment, so as to obtain more comprehensive understanding of dipolar active systems.

\section*{Conflicts of interest}

There are no conflicts to declare.

\appendix
\section{Appendix}

\subsection{\label{sec:Ewald}Ewald summation}

To deal with the long-range character of the dipole-dipole interactions, we employ 2D Ewald summation with the ``tinfoil'' boundary condition. \cite{Mazars2011} Within this method, the total dipole-dipole energy is separated into different contributions,

\begin{align} \label{eqn:Ewald-potential}
  U_{dd} =
         & \dfrac{1}{2} \sum_{i \neq j}^{N}
             \Bigg[
               \big(
                 \boldsymbol{\mu}_{i} 
                 \cdot 
                 \boldsymbol{\mu}_{j}
               \big)
               B\left(r_{ij}, \alpha\right)
               -
               \big(\boldsymbol{\mu}_{i} \cdot \boldsymbol{r}_{ij} \big)
               \big(\boldsymbol{\mu}_{j} \cdot \boldsymbol{r}_{ij} \big)
               C\big(r_{ij}, \alpha\big)
             \Bigg] 
           \nonumber \\
         & +
           \dfrac{1}{L^2} \sum_{\boldsymbol{k} \neq 0} 
           \dfrac{
             \pi
           }{
             k
           }
           \text{erfc}
             \left(
               \dfrac{k}{2\alpha}
             \right)
           \vert
             \mathzapf{M}\left(\boldsymbol{k}\right)
           \vert^2
           \nonumber \\
         & -
           \dfrac{2\alpha^3}{3\sqrt{\pi}}N\mu^2
           \text{.}
\end{align}
The first term on the right-hand side of eqn~\eqref{eqn:Ewald-potential} corresponds to the real-space contribution, where the functions $B\left(r, \alpha\right)$ and $C\left(r, \alpha\right)$ are defined by \cite{Klapp2002}
\begin{align}
B\left(r, \alpha\right) \equiv 
  &
  \dfrac{1}{r^3}
  \bigg[
    \dfrac{2 \alpha r}{\sqrt{\pi}}
    \text{exp}\left(-\alpha^2 r^2 \right)
    +
    \text{erfc}\left(\alpha r\right)
  \bigg] \text{, and} \\
C\left(r, \alpha\right) \equiv
  &
  \dfrac{1}{r^5}
  \bigg[
    \dfrac{2 \alpha r}{\sqrt{\pi}}
    \left(
      3 + 2 \alpha^2 r^2
    \right)
    \text{exp}\left(-\alpha^2 r^2 \right)
    +
    3 \: \text{erfc}\left(\alpha r\right)
  \bigg] \label{eqn:CoefC}
\end{align} 
with erfc($x$) being the complementary error function.
In eqn~\eqref{eqn:Ewald-potential}, the real-space contribution is formulated under the assumption that the convergence parameter $\alpha$ is large enough, such that we can consider only the interactions within the central simulation box.
This is achieved by choosing $\alpha = 7/L$ (with $L$ being the box size) and evaluating $\boldsymbol{r}_{ij}$ with the minimum-image convention. \cite{Klapp2002}
The second term in eqn~\eqref{eqn:Ewald-potential} represents the reciprocal-space contribution with
\begin{align}
  \mathzapf{M}\left(\boldsymbol{k}\right) 
    & = 
    \sum_{j=1}^N 
      \big(
      \boldsymbol{k} \cdot \boldsymbol{\mu}_j
      \big)
      \text{exp}
      \big(
        - \mathzapf{i} \boldsymbol{k} \cdot \boldsymbol{r}_j
      \big) \nonumber \\
    & \equiv
    \text{Re}
    \left\lbrace
      \mathzapf{M}
    \right\rbrace
    +
    \mathzapf{i}\;
    \text{Im}
    \left\lbrace
      \mathzapf{M}
    \right\rbrace
      \text{.} \label{eqn:FourierM}
\end{align}
The wave vectors in the reciprocal (square) lattice are given by $\boldsymbol{k} = \left( 2 \pi / L \right) \boldsymbol{m}$, where $\boldsymbol{m} = \left(m_x, m_y\right)^T$ with $m_x$ and $m_y$ being integers.
The magnitude of the wave vector is denoted as $k = 2\pi\sqrt{m_x^2 + m_y^2}/L$.
We evaluate the reciprocal-space summation for the $\boldsymbol{k}$-vectors within the range $\boldsymbol{m}^2 = m_x^2 + m_y^2 \leq 15^2$. \cite{Schoen2007}
The third term represents the correction due to unphysical self-interaction of dipole moments. \cite{Schoen2007}
%

%
The Langevin equations~\eqref{eqn:coupled_Langevin_trans} and~\eqref{eqn:coupled_Langevin_rot} involve the forces and torques due to the dipolar interactions.
From eqn~\eqref{eqn:Ewald-potential}, the force acting on the $i$th particle due to dipole-dipole interactions is given by
\begin{align}
  \boldsymbol{F}_{i, dd} & = - \nabla_{\boldsymbol{r}_{i}} U_{dd} \nonumber \\
                         & = \boldsymbol{F}_{i, dd}^{R} + \boldsymbol{F}_{i, dd}^{\boldsymbol{k} \neq 0}\text{,}
\end{align}
where the real-space contribution is 
\begin{align}
  \boldsymbol{F}_{i, dd}^{R} = 
  \sum_{i \neq j} 
    \bigg[ &
        \Big(
        \big(
          \boldsymbol{\mu}_{i} \cdot \boldsymbol{\mu}_{j}
        \big)
        \boldsymbol{r}_{ij}
        +
        \boldsymbol{\mu}_{i}
        \big(
          \boldsymbol{\mu}_{j} \cdot \boldsymbol{r}_{ij}
        \big)
        +
        \boldsymbol{\mu}_{j}
        \big(
          \boldsymbol{\mu}_{i} \cdot \boldsymbol{r}_{ij}
        \big)
      \Big)
      C\left(r_{ij}, \alpha \right) \nonumber \\
    & -
      \big(
          \boldsymbol{\mu}_{i} \cdot \boldsymbol{r}_{ij}
      \big)
      \big(
          \boldsymbol{\mu}_{j} \cdot \boldsymbol{r}_{ij}
      \big)
      \boldsymbol{r}_{ij}
      D\left(r_{ij}, \alpha \right)
    \bigg]\text{.}
\end{align}
The function $C\left(r,\alpha\right)$ is defined in eqn~\eqref{eqn:CoefC}, and $D\left(r,\alpha\right)$ is given by
\begin{align}
D\left(r, \alpha\right) \equiv
    \dfrac{1}{r^7}
    \bigg[ &
      \dfrac{2 \alpha r}{\sqrt{\pi}}
      \left(
        15 + 10 \alpha^2 r^2 + 4 \alpha^4 r^4
      \right)
      \text{exp}\left(-\alpha^2 r^2 \right) \nonumber \\
      & + 
      15 \; \text{erfc}
      \left(
        \alpha r
      \right)
    \bigg]\text{.}
\end{align}
Further, the contribution from the Fourier-form is written as
\begin{align}
\boldsymbol{F}_{i, dd}^{\boldsymbol{k} \neq 0} = 
  \dfrac{2 \pi}{L^2}
  \sum_{\boldsymbol{k} \neq 0} &
  \dfrac{
    \boldsymbol{k}
    \left(
      \boldsymbol{k}
      \cdot
      \boldsymbol{\mu_i}
    \right)    
  }
  {
    k
  }
  \text{erfc}
  \left(
    \dfrac{k}{2\alpha}
  \right) \nonumber \\
  & \Big(
    \text{sin}
    \left(
      \boldsymbol{k}
      \cdot
      \boldsymbol{r}_{i}
    \right)
    \text{Re}
    \big\lbrace
      \mathzapf{M}\left(\boldsymbol{k}\right)
    \big\rbrace
    +
    \text{cos}
    \left(
      \boldsymbol{k}
      \cdot
      \boldsymbol{r}_{i}
    \right)
    \text{Im}
    \big\lbrace
      \mathzapf{M}\left(\boldsymbol{k}\right)
    \big\rbrace
  \Big)\text{,}
\end{align}
where $\text{Re}\left\lbrace\mathzapf{M}\left(\boldsymbol{k}\right)\right\rbrace = \sum_{j=1}^N 
      \big(
      \boldsymbol{k} \cdot \boldsymbol{\mu}_j
      \big)
      \text{cos}
      \left(
        \boldsymbol{k} \cdot \boldsymbol{r}_j
      \right)$ and 
      $\text{Im}\left\lbrace\mathzapf{M}\left(\boldsymbol{k}\right)\right\rbrace = -\sum_{j=1}^N 
      \big(
      \boldsymbol{k} \cdot \boldsymbol{\mu}_j
      \big)
      \text{sin}
      \left(
        \boldsymbol{k} \cdot \boldsymbol{r}_j
      \right)$ [see eqn~\eqref{eqn:FourierM}].
%

%
In the present work, every dipole moment is restricted to point in the $xy$-plane with $\psi_{i}$ characterizing its direction relative to the $x$-axis. 
Therefore, the torque acting on the $i$th particle is always directed along the $z$-axis.
It is given by
\begin{align}
\boldsymbol{T}_{i, dd} & = - \partial_{\psi_{i}} U_{dd} \;\hat{\boldsymbol{z}} \nonumber \\
                       & = \boldsymbol{T}_{i}^{R}+\boldsymbol{T}_{i}^{\boldsymbol{k} \neq 0}\text{,}
\end{align}
where $U_{dd}$ is given in eqn~\eqref{eqn:Ewald-potential}, and $\hat{\boldsymbol{z}}$ denotes the unit vector along the positive $z$-axis.
The real-space component of the torque is given by 
\begin{equation}
  \boldsymbol{T}_{i}^{R} = 
  -
  \sum_{j \neq i}
  \bigg[
    \big(
      \boldsymbol{\mu}_{i} \times \boldsymbol{\mu}_{j}
    \big)
    B\left( r_{ij}, \alpha \right)
    -    
    \big(
      \boldsymbol{\mu}_{i} \times \boldsymbol{r}_{ij}
    \big)
    \big(
      \boldsymbol{\mu}_{j} \cdot \boldsymbol{r}_{ij}
    \big)
    C\left( r_{ij}, \alpha \right)
  \bigg]\text{.}
\end{equation}
Finally, the reciprocal-space contribution is given by
\begin{align}
  \boldsymbol{T}_{i}^{\boldsymbol{k} \neq 0} = 
  \dfrac{2 \pi}{L^2}
  \sum_{\boldsymbol{k} \neq 0} &
  \dfrac{
    \boldsymbol{k} \times \boldsymbol{\mu}_{i}
  }
  {
    k
  }
  \;
  \text{erfc}
  \left(
    \dfrac{k}{2\alpha}
  \right) \nonumber \\
  & \Big(
    \text{cos}
    \left(
      \boldsymbol{k} \cdot \boldsymbol{r}_{i}
    \right)
    \text{Re}
    \big\lbrace
      \mathzapf{M}\left(\boldsymbol{k}\right)
    \big\rbrace
    -
    \text{sin}
    \left(
      \boldsymbol{k}
      \cdot
      \boldsymbol{r}_{i}
    \right)
    \text{Im}
    \big\lbrace
      \mathzapf{M}\left(\boldsymbol{k}\right)
    \big\rbrace
  \Big)\text{.}
\end{align}

\subsection{\label{sec:clusterDistri}The cluster size distribution}

\begin{figure}[!htbp]
  \centering
  \includegraphics[width=0.9\linewidth]{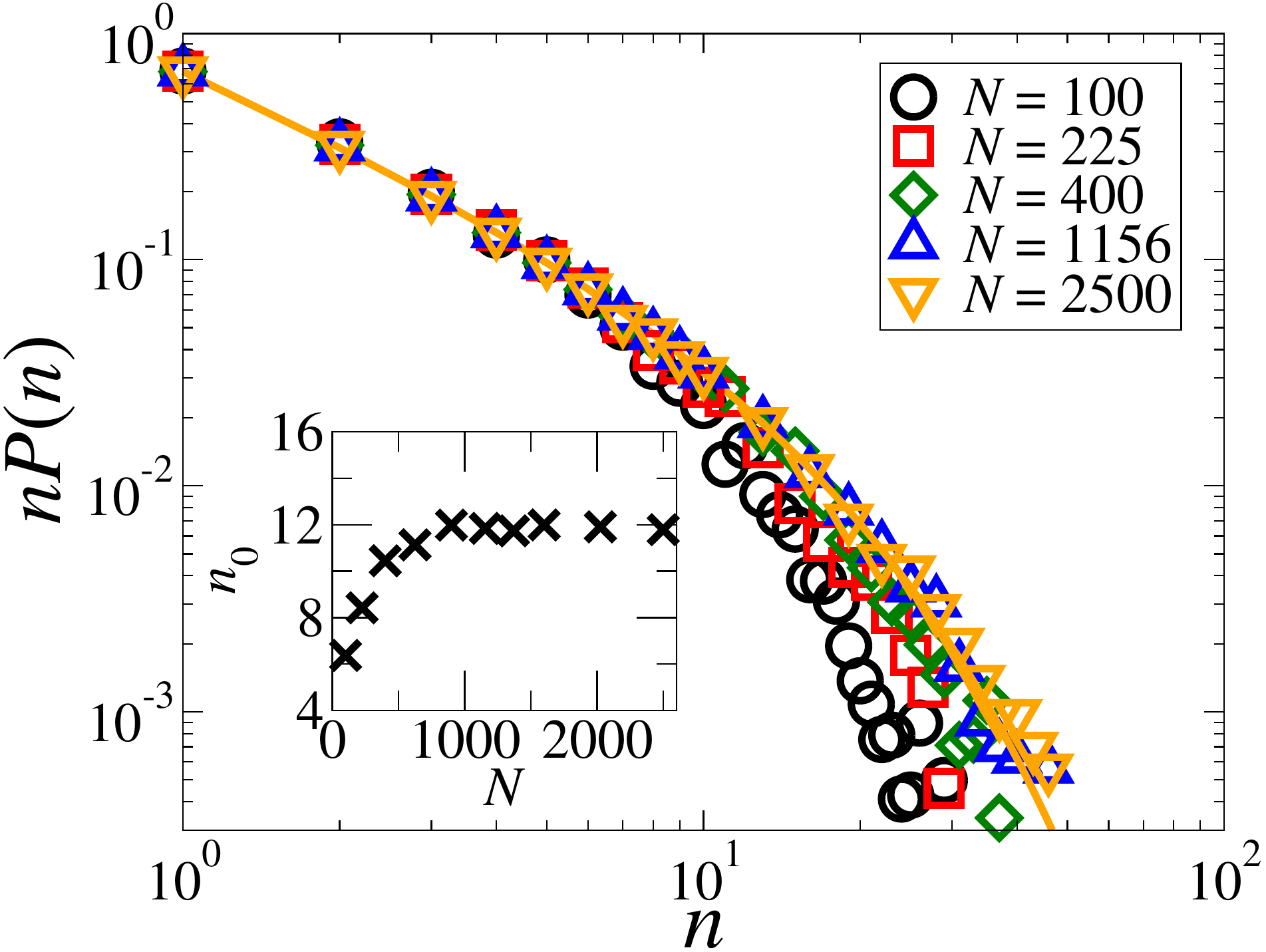}
\caption[The cluster size distribution at $\Phi = 0.23$]{(Color online) Weighted distribution of cluster size at $\Phi = 0.23$ for the particle number $N = 100$ (black circles), $225$ (red squares), $400$ (green diamonds), $1156$ (blue triangles up), and $2500$ (orange triangles down). The orange solid line indicates the curve fitted to $N = 2500$ according to eqn~\ref{eqn:clusterDistri}. Inset: Characteristic size of clusters, $n_0$, as a function of particle number $N$.}
\label{fig:P02clusterDistri}
\end{figure}

%
The purpose of this paragraph is to show that the clusters observed in the ``micro-flocking'' state at intermediate densities are indeed microscopic structures.
To this end, we performed a finite size analysis of the cluster size distribution.
If these clusters are microscopic patterns, they should not grow as the total number of particles $N$ (\textit{i.e.}, the system size) becomes larger.
Figure~\ref{fig:P02clusterDistri} shows the weighted cluster size distribution $nP\left(n\right)$ at $\Phi = 0.23$, $v_0^* = 100$ and $\lambda = 6$ for various system sizes $N$, where the cluster size $n$ represents the number of particles within a cluster.
For $N < 1000$, the distribution decays faster as $N$ decreases, indicating that the systems are influenced by the finite sizes.
However, as we increase the particle number up to $N \gtrsim 1000$, the data points collapse onto a single curve.
To obtain a more quantitative description, we fit the data points for each system size via the function,
\begin{equation} \label{eqn:clusterDistri}
  nP\left(n\right) = P\left(1\right)n^{-1}\textit{e}^{-\left(n-1\right)/n_0}\text{,}
\end{equation}
where $n_0$ denotes the characteristic cluster size.
Similar fitting functions of the cluster size distribution have also been considered in systems of conventional active Brownian particles \cite{Fily2014} and polar active disks. \cite{Martin-Gomez2018}
The inset of Fig.~\ref{fig:P02clusterDistri} shows that $n_0$ reaches a plateau once $N \gtrsim 1000$.
This suggests that the cluster size indeed remains constant as the system size $N$ increases.
By this, we can confirm that the clusters observed in the ``micro-flocking'' state at $\Phi = 0.23$ are microscopic structures.
%

\begin{figure}[!htbp]
  \centering
  \includegraphics[width=0.9\linewidth]{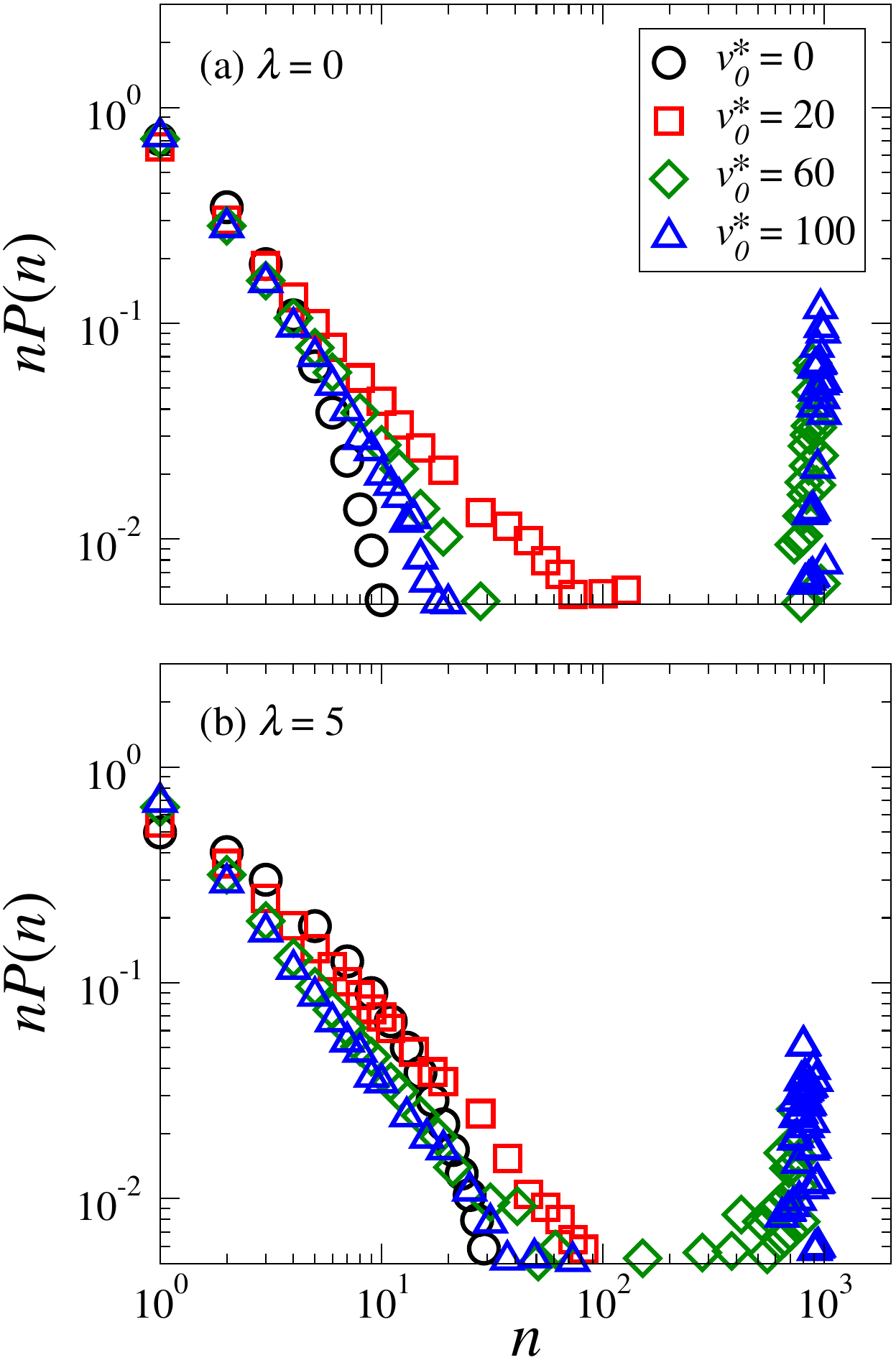}
  \caption[The weighted distribution function of cluster size at $\Phi = 0.58$]{(Color online) Weighted cluster size distribution at $\Phi = 0.58$ for the motility $v_0^* = 0$ (black circles), $20$ (red squares), $60$ (green diamonds), and $100$ (blue triangles) with the coupling strength $\lambda = 0$ (a) and $\lambda = 5$ (b). We denote $n$ as the cluster size.}
  \label{fig:clusterDistri}
\end{figure}

%
To provide additional information on the clustering behavior at high densities ($\Phi = 0.58$), we plot in Fig.~\ref{fig:clusterDistri} the weighted cluster size distribution for non-dipolar and strong coupled dipolar active particles at $\Phi = 0.58$.
At the coupling strength $\lambda = 0$, our model reduces to the limiting case of active Brownian particles. \cite{Stenhammar2013, Buttinoni2013}
Here, the correspondent weighted cluster size distribution function undergoes a transition from an exponential decay ($v_0^* \lesssim 20$) into a curve with a power law decay at small cluster sizes $n$ and a peak at large $n$ ($v_0^* \gtrsim 20$), as shown in Fig.~\ref{fig:clusterDistri}(a).
Interestingly, the distribution functions of strongly-coupled dipolar particles ($\lambda = 5$) bear resemblance to the case of active Brownian particles, as can be seen in Fig.~\ref{fig:clusterDistri}(b).
For passive dipolar particles ($v_0^* = 0$), the weighted distribution function vanishes at $n \approx 30$, which is three times as large as in the non-dipolar case.
This is due to the fact that the strongly-coupled particles tend to form head-to-tail configurations and therefore display chain-like structures.
With increasing motilities, such structures disappear.
Further, the dipolar particles start to align with their neighbors and form into clusters, as shown in Fig.~\ref{fig:P05Snapshots}(d).
As indicated by the peaks at $n \approx 10^3$ in Fig.~\ref{fig:clusterDistri}(b), these particles form ``giant'' clusters when $v_0^* \gtrsim 20$.

\section*{Acknowledgements}

The authors would like to thank Deutsche Forschungsgemeinschaft for the financial support from GRK 1524 (DFG No. 599982).
This work was also supported by NSF's Research Triangle MRSEC under grant number DMR 1121107.




\renewcommand\refname{Notes and references}

\bibliography{DABPs} 

\begin{mcitethebibliography}{86}
\bibitem[phi()]{phi_alternative}
{The alternative definition of the area fraction is given by $\tilde{\Phi} = N
  \pi \sigma^2 /\left(4L^2\right)$. Therefore, $\Phi = \tilde{\Phi}
  \left(\sigma_{eff}/\sigma\right)^2$. In this work, we consider three
  different density values $\tilde{\Phi} = 0.1$, $0.2$, and $0.5$.
  Consequently, the exact area fractions to the fifth decimal place are $\Phi =
  0.11632$, $0.23264$, and $0.58160$.}\relax
\mciteBstWouldAddEndPunctfalse
\mciteSetBstMidEndSepPunct{\mcitedefaultmidpunct}
{}{\mcitedefaultseppunct}\relax
\EndOfBibitem
\bibitem[rL()]{rL}
{The pair correlation function has been obtained with a spatial resolution
  $\Delta r = 0.0125 \sigma$. The resulting exact values for $r_{L}$ are $r_{L}
  = 1.15625 \sigma$ for $\Phi = 0.1$, $r_{L} = 1.13125 \sigma$ for $\Phi =
  0.3$, and $r_{L} = 1.11875 \sigma$ for $\Phi = 0.5$.}\relax
\mciteBstWouldAddEndPunctfalse
\mciteSetBstMidEndSepPunct{\mcitedefaultmidpunct}
{}{\mcitedefaultseppunct}\relax
\EndOfBibitem
\end{mcitethebibliography}
\bibliographystyle{rsc} 

\end{document}